\documentclass[aps,prb,twocolumn,superscriptaddress,a4paper]{revtex4-2}
\usepackage[dvipdfmx]{graphicx}
\usepackage{mathrsfs}
\usepackage{amsmath,amsthm,amssymb}
\usepackage{amsmath,bm}
\usepackage{color}
\usepackage{ifthen}
\usepackage[svgnames,psnames]{xcolor}
\usepackage[colorlinks,citecolor=blue,linkcolor=red,urlcolor=blue]{hyperref}
\usepackage{physics}
\usepackage{mathtools}
\usepackage{multirow,bigdelim}
\usepackage{float}
\usepackage{enumerate}

\def\ket#1{|#1\rangle }
\def\bra#1{\langle #1 |}

\def\Tr{\mathrm{Tr}}

\newcommand{\last}[1]{\textcolor{black}{#1}}
\newcommand{\rev}[1]{\textcolor{black}{#1}}
\newcommand{\res}[1]{\textcolor{black}{#1}}

\newcommand{\refirst}[1]{\textcolor{black}{#1}}

\newcommand{\sgn}{\mathrm{sgn}\!}

\newlength{\figwidth}
\newlength{\figlarge}
\begin{document}
\title{
Interaction-induced Liouvillian skin effect in a fermionic chain with a two-body loss
}
\author{
Shu Hamanaka
}
\email{hamanaka.shu.45p@st.kyoto-u.ac.jp}
\affiliation{
Department of Physics, Kyoto University, Kyoto 606-8502, Japan
}

\author{
Kazuki Yamamoto
}
\affiliation{
  Department of Physics, Tokyo Institute of Technology, Meguro, Tokyo 152-8551, Japan
}
\affiliation{
Department of Physics, Kyoto University, Kyoto 606-8502, Japan
}

\author{
Tsuneya Yoshida
}
\affiliation{
Department of Physics, Kyoto University, Kyoto 606-8502, Japan
}

\date{\today}
\begin{abstract}
\rev{
Despite recent intensive research on topological aspects of open quantum systems, effects of strong interactions have not been sufficiently explored. In this paper, we demonstrate that \refirst{complex-valued} interactions induce the Liouvillian skin effect by analyzing a one-dimensional correlated model with two-body loss. \res{We show that}, in the presence of \refirst{complex-valued} interactions, eigenmodes and eigenvalues of the Liouvillian strongly depend on boundary conditions. \res{Specifically, we find that} \refirst{complex-valued} interactions induce localization of eigenmodes of the Liouvillian around the right edge under open boundary conditions.
\res{To characterize the Liouvllian skin effect, we define the topological invariant by using the Liouvillian superoperator.
Then, we numerically confirm that the topological invariant captures the Liouvillian skin effect.} 
Furthermore, the presence of the localization of eigenmodes results in the unique dynamics observed only under open boundary conditions: particle accumulation at the right edge in transient dynamics.
Our result paves the way to \res{realize topological phenomena in open quantum systems induced by strong interactions.}}

\end{abstract}
\maketitle


\section{Introduction}
In the past decade, \res{a lot of theoretical and experimental studies have}
uncovered topological aspects of condensed matter systems~\cite{Kane-Mele1,Kane-Mele2,TQFT,Zhang2009,Hasan-Kane,Qi-Zhang,Schnyder,Ryu-2010,Kitaev-periodic,Sato_2017}.
\rev{In particular, it has been elucidated that strong correlations alter topological phases and lead to novel phenomena.
For example, it has been turned out that interactions change the $\mathbb{Z}$-classification to the $\mathbb{Z}_{8}$-classification for one-dimensional topological superconductors~\cite{Fidkowski-Kitaev-1, Fidkowski-Kitaev-2}
and another study has shown that strong correlations generate topological Mott phases~\cite{TopoMot}.}
\rev{Moreover, \res{in Ref.~\cite{Enabled}, the interaction-enabled topological insulator has been proposed, which has no counterpart in noninteracting systems.}}

On the other hand, \rev{non-Hermitian physics has attracted broad interest in classical and open quantum systems
~\cite{Gong-Adv,Gong-PRX,Kawabata-PRX,Esaki_2011,Biorthogonal,
Bergholtz-Review, Okuma-Review, Yokomizo-Murakami, Shen-Zhen-Fu, Lee-Review, Kozii-Fu, Yoshida-Ring, Yoshida-Heavy, Yao-Fei-Wang, Budich, Okugawa, Edvardsson, Knotted,Borgnia}.
One of the most remarkable phenomena induced by non-Hermiticity is the non-Hermitian skin effect, \res{which is characterized by} the extreme sensitivity of eigenvalues and eigenstates to boundary conditions~\cite{Yao-Wang, Lee, Okuma-Topo, Zhang-Topo, Kawabata-Higher, Song-PRL-2019, Li2020, Longhi, Kawabata-entangle, Okugawa-skin, Yoshida-Mirror}.} 
\rev{The non-Hermitian skin effect has been experimentally observed 
in \res{ultracold $^{87}$Rb atoms}~\cite{Liang-PRL-2022} as well as electric circuits~\cite{Hofmann-2020}, quantum walks~\cite{Xiao-2020}, and mechanical metamaterials~\cite{Ghatak-2020}.
\rev{\res{In noninteracting systems}}, theoretical studies have shown that the non-Hermitian skin effect is caused by the nontrivial point-gap topology, which is \rev{intrinsic} to non-Hermitian systems~\cite{Lee, Okuma-Topo, Zhang-Topo}.
Furthermore, the non-Hermitian skin effect has been extended to open quantum systems following the Lindblad master equation~\cite{Yang-2022,Haga-PRL,Pan-PRB-2021,Song-PRL-2019,Liu-PRR-2020,Mori-PRL}.
\res{Especially, the Liouvillian skin effect manifests as the extreme dependence of eigenvalues and eigenmodes of the Liouvillian on boundary conditions.
In particular, the eigenmode localized near the edge is referred to as the skin mode}.
}
\res{It has been pointed out that} the Liouvillian skin effect has a striking influence on the relaxation processes. 
Specifically, it has been reported that the maximal relaxation time \rev{to the steady state} can diverge while maintaining the Liouvillian gap finite~\cite{Haga-PRL, Mori-PRL}.

\last{
In addition to the above progress of the non-Hermitian topological band theory,
it has become possible to implement dissipative correlated systems in ultracold atoms~\cite{Honda, Tomita1, Tomita2, Syassen, Barontini, Labouvie, Sponselee, Bouganne, Yan2013}.
This development has opened up a new direction in studies of novel phases and phenomena, such as nonequilibrium steady states~\cite{Yamamoto-PRR-2020, Yoshioka} and dynamical phase transitons~\cite{incoherenton, lindbladian}.}
Previous studies have revealed that particle losses induce unique phenomena. 
In particular, two-body loss brings about unusual behavior, e.g., the sign reversal 
of magnetic correlations~\cite{Honda, Nakagawa-PRL-2020}.
Moreover, a lot of theoretical studies have been conducted on a variety of quantum many-body phenomena with atom losses~\cite{Yamamoto-PRB1, Yamamoto-PRB2, Rosso1, Rosso2, Zeno, Ripoll, Ashida-PRA, Perfetto-PRL-2023, Perfetto-arxiv-2023}, such as unconventional superfluid phase transitions in a dissipative BCS model~\cite{Yamamoto-PRL,Yamamoto3} and anomalous dissipation-induced renormalization-group flows in a non-Hermitian Kondo model~\cite{Nakagawa-Kondo}.

\rev{
In view of the pivotal role of interactions in enriching topological phases in Hermitian systems and inducing unique phenomena in open quantum systems,
one may naturally expect the presence of novel phenomena induced by the interplay between strong interactions and non-Hermitian topology.
So far, the effects of interactions of non-Hermitian topological phases have been studied in several works
~\cite{Yoshida-interactingEP-PRB-2023,Yoshida2019, Yoshida-Reduction1-PRB-2022,Yoshida-Reduction2-PRB-2021,Kawabata-Manybody-PRB-2022, Faugno-PRL-2022,Zhang-Song-PRB-2022,Taozhi-2023,wang-2023,mao-2022,Kawabata-entangle,micallo-2023,Luitz-2019,Liu-2020,Lee-PRB-2001,Shen-2022,Alsallom,Tsubota, Can-arxiv-2023}.}
\res{However, previous studies have mainly focused on the effective Hamiltonian, which captures the time evolution of a single trajectory between successive quantum jumps~\cite{Daley}.} 
\rev{\res{Thus,} it seems that the effects of interaction on the topological property of the Liouvillian remain unclear~\cite{Lieu-PRL-2020, Yoshida-fractional-PRR-2020, Kawasaki-PRB-2022, Kawabata-Liouvillian-symmetry-2022,li-2023}.
More specifically, \res{whether} many-body interactions can induce the Liouvillian skin effect has not been addressed.}

\refirst{
In this work, we demonstrate that complex-valued interaction can induce the Liouvillian skin effect in one-dimensional open quantum systems.}
Specifically, we analyze the correlated fermionic systems with two-body loss.
\res{We show that} owing to strong interactions, eigenmodes and eigenvalues of the Liouvillian become extremely sensitive to boundary conditions.
In particular, eigenmodes of the Liouvillian exhibit localization near the edge.
\rev{To characterize the Liouvillian skin effect, we introduce the topological invariant defined by the Liouvillian superoperator.
Then, we numerically reveal that the above topological invariant characterizes the Liouvillian skin effect.}
\res{Moreover,} the Liouvillian skin effect significantly affects the dynamics. 
In particular, in transient dynamics, particles accumulate near the right edge under open boundary conditions (OBC).

The rest of this paper is organized as follows.
In Sec.~\ref{sec:model}, \res{we first}
introduce the dissipative one-dimensional correlated model.
We then briefly explain the methods to analyze the Lindblad equation
via the vectorization of the density matrix.
\rev{Section~\ref{sec:definition} provides the definition of the topological invariant and \res{the right-state particle density, which measures the degree of localization of eigenmodes in many-body systems.}}
Then, in Sec.~\ref{sec:result}, a numerical demonstration of the interaction-induced Liouvillian skin effect is conducted.
We give the \rev{conclusions} in Sec.~\ref{sec:discussions}.
In Appendix~\ref{app:trivial-winding}, we discuss the relation between the  symmetry of the Liouvillian and the topological number.
In Appendix~\ref{app:Haga}, we compute the topological number analytically and give the characterization of the Liouvillian skin effect reported in Ref~\cite{Haga-PRL}.
We numerically show the absence of the Liouvillian skin effect in noninteracting systems in Appendix~\ref{app:no-dissipation}.
Appendix~\ref{app:dep} is devoted to the sensitivity of eigenvalues
of the Liouvillian to boundary conditions.
We provide the derivation of an alternative method for calculating the topological number in Appendix~\ref{app:relationproof}.
\refirst{Appendix~\ref{app:two-step} gives the results about slowing down process corresponding to the Liouvillian skin effect.}
In Appendix~\ref{app:other-case}, we demonstrate that the Liouvillian skin effect survives for other configurations of down-spins.
\refirst{In Appendix~\ref{app:deform}, we give the reason why particles are localized near the right edge.}

\section{Model and Method}
\label{sec:model}
\subsection{Falicov-Kimball model with two-body loss}
We consider the two-orbital Falicov-Kimball model~\cite{FK}
\begin{align}
    \label{eq:Hamiltonian}
    H &= \sum\limits_{\langle ij \rangle \alpha \beta } h_{ i\alpha j\beta}
    {c}^{\dagger}_{i\alpha\uparrow}{c}_{j\beta\uparrow}
    + U \sum\limits_{j} {n}_{j b \uparrow} {n}_{j b \downarrow},
\end{align}
where ${c}^{\dagger}_{j \alpha\sigma}({c}_{j\alpha\sigma})$ is
a fermionic creation (annihilation) operator at site $j=1,\cdots,L$ in orbital $\alpha=a,b$ with  spin $\sigma=\uparrow, \downarrow$ state~\footnote{We note that hopping of down-spin states excluded, which simplifies the computation of eigenstates and eigenvalues of $H_{\mathrm{eff}}$ [see Eq.~\eqref{eq:eff-Hamil}].
Introducing the hopping of down-spin states may change the eigenvalues. We note, however, that the topological properties remain unchanged as long as the point-gap opens.}.
$h_{i\alpha j\beta}$ is the hopping Hamiltonian between site $i$ in orbital $\alpha$ with spin up state and site $j$ in orbital $\beta$ with spin up state.
$U$ denotes the strength of interactions.
The summation of the first term $\langle ij \rangle$ runs over all pairs of nearest neighbor sites $i$ and $j$.
By applying the Fourier transformation to the first term in Eq.~(\ref{eq:Hamiltonian}), the Bloch Hamiltonian $h_{\alpha\beta}(k)$ in the orbital space reads 
\begin{align}
    h(k) = b_{2}(k){\sigma}_{2}+b_{3}(k){\sigma}_{3},
    \label{eq:Bloch-Hamiltonian}
\end{align}
with
\begin{subequations}
    \begin{align}
         & b_{2} =  2t_{h}-0.5t_{h} \sin k, \label{eq:b2} \\
         & b_{3} =  2t_{h} \cos k. \label{eq:b3}
    \end{align}
\end{subequations}
Here, $\sigma_{j}$ $(j=1,2,3)$ express the Pauli matrices in the orbital space.
\res{The Hamiltonian given in Eq.~(\ref{eq:Hamiltonian}) is obtained from the two-orbital Hubbard model by turning off the hopping of fermions in the down-spin states.
}
\refirst{
It is worth noting that the above model breaks the inversion symmetry. Breaking the inversion symmetry is essential for the emergence of the interaction-induced Liouvillian skin effect  (see Appendix~\ref{app:trivial-winding}), rather than the specific values of the specific value  $2t_h,-0.5t_h$ and $2t_h$ in Eqs.~\eqref{eq:b2} and \eqref{eq:b3}. We also note that the multibandness is essential to inducing the skin effect without asymmetric hopping.}

When dissipation is introduced into this model, under the Markov approximation, the dynamics is described by the Lindblad equation~\cite{Lindblad, Gorini}
\begin{align}
    \label{eq:Lindblad}
    \frac{d{\rho}}{dt} = \mathscr{L}(\rho)
    =
    -i[H,\rho] + \sum\limits_{j}\biggl[
        L_{j}\rho L^{\dagger}_{j}
    -
    \frac{1}{2}\{L_{j}^{\dagger}L_{j},\rho\}
    \biggr].
\end{align}
Here, $\mathscr{L}$ denotes the Liouvillian, \res{which is} the superoperator acting on
the density matrix ${\rho}$, the operator
$H$ is the Hamiltonian, and the Lindblad operator $L_{j}$
characterizes the effect of dissipation.
The Lindblad operator is given by the on-site two-body loss
\begin{align}
\label{eq:dissipator}
    {L}_{j} =  \sqrt{2 \gamma} {c}_{jb\uparrow}{c}_{jb\downarrow}.
\end{align}
\res{
We decompose the Liouvillian  $\mathscr{L}(\rho)$ as
\begin{align}
    \mathscr{L}(\rho) = \mathscr{L}_{0}(\rho)+\mathscr{L}_{\mathrm{J}}(\rho),
\end{align}
where we have introduced
}
\begin{align}
    \label{eq:L0}
    \mathscr{L}_{0} (\rho) = -i({H}_{\rm{eff}}{\rho}-{\rho} {H}_{\rm{eff}}^{\dagger})  
\end{align}
and
\begin{align}
    \label{eq:LJ}
    \mathscr{L}_{\mathrm{J}} (\rho) =  \sum\limits_{j} {L}_{j}{\rho} {L}_{j}^{\dagger}.  
\end{align}
Here, the non-Hermitian Hamiltonian given by
\begin{align}
\label{eq:eff-Hamil}
    H_{\rm{eff}} &= H-\frac{i}{2}\sum_{j}L^{\dagger}_{j}L_{j}  \nonumber \\
    &=  \sum\limits_{\langle ij \rangle \alpha \beta } h_{ i\alpha j\beta}
    {c}^{\dagger}_{i\alpha\uparrow}{c}_{j\beta\uparrow}
    + (U-i\gamma) \sum\limits_{j} {n}_{j b \uparrow} {n}_{j b \downarrow}
\end{align} 
describes the dynamics of the single quantum trajectory between the quantum jumps~\cite{Daley}.
\refirst{In the following, we demonstrate that the complex-valued interaction $U-i\gamma$ induces the Liouvillian skin effect~\footnote{The orbital dependence of the anti-Hermitian part is essential for the interaction induced Liouvillian skin effect. As a simple example, we introduce interaction $U$ and $\gamma$ only to orbital $b$. We consider that introducing complex-valued interaction does not change the topology as long as the point-gap opens. We also note that by fixing the configuration of down-spins, the effective Hamiltonian takes a quadratic form in the subspace.}.}

\subsection{Vectorization of the density matrix}
\rev{
In this subsection, we rewrite the Liouvillian superoperator $\mathscr{L}$
 as an operator $\mathcal{L}$ acting on the doubled Hilbert space by vectorizing the density matrix.}
Following the procedure of Refs.\cite{Shibata,Yoshioka,incoherenton,lindbladian},
we identify the density matrix $\rho$ as a vector $\ket{\rho}\rangle$  in the doubled Hilbert space $\mathcal{H}\otimes\mathcal{H}$ through the mapping
\begin{align}
\label{eq:map}
   \rho =  \sum\limits_{ij} 
   \rho_{ij} \ket{i} \bra{j}
    \mapsto \ket{\rho}\rangle =  \sum\limits_{ij} 
    \rho_{ij} \ket{i} \otimes \ket{{j}}.
\end{align} 
We note that the first (second) space of the doubled Hilbert space $\mathcal{H}\otimes\mathcal{H}$ is referred to as the ket (bra) space.
When the density matrix $\rho$ is given by the vectorized form $\ket{\rho}\rangle$, 
the Liouvillian superoperator $\mathscr{L}$ is written as the operator $\mathcal{L}$ that acts on the doubled Hilbert space 
\begin{align}
    \label{eq:Liouvillian}
    \mathcal{L}= \mathcal{L}_{0} + \mathcal{L}_{\mathrm{J}}.
\end{align}
Here, we define
\begin{align}
\mathcal{L}_{0} = -i\Bigl( H_{\rm{eff}}\otimes {I} - I \otimes {H}_{\rm{eff}}^{*}\Bigr)
\end{align}
and 
\begin{align}
\label{eq:jump}
    \mathcal{L}_{\mathrm{J}} = \sum\limits_{j} L_{j} \otimes {L}_{j}^{*},
\end{align}
where $I$ is the identity operator acting on the ket or bra space
\footnote{We note that $c$'s satisfy the anticommutation relation each other as well as  $\tilde{c}$'s. Here, we denote the operator marked with a tilde as acting on the bra space. However, $c_{j\alpha\sigma}$ and $\tilde{c}_{j'\alpha'\sigma'}$ commute $[c_{j\alpha\sigma}, \tilde{c}_{j'\alpha'\sigma'}]=0$.
We note however, that the following results can be obtained  by introducing $d_{j\alpha\sigma}=c_{j\alpha\sigma}$ and ${d}_{j'\alpha'\sigma'}=P_{\mathrm{f}\mathrm{K}}\tilde{c}_{j'\alpha'\sigma'}$ with $P_{\mathrm{f}\mathrm{K}}=e^{i\pi \sum_{j\alpha\sigma} c^{\dagger}_{j\alpha\sigma} c_{j\alpha\sigma}}$.}.
Thus, the Liouvillian superoperator $\mathscr{L}$ is mapped to the non-Hermitian operator $\mathcal{L}$ acting on the doubled Hilbert space.
\res{After the vectorization of the density matrix, the $n$-th eigenmode $\ket{{\rho}^{(n)}_{\mathrm{R}}}\rangle$ and the
 $n$-th eigenvalue  $\Lambda_{n}$ are obtained by solving the eigenvalue equation}
\begin{align}
    \label{eq:eigenvalue-Liouvillian}
    \mathcal{L}\ket{{\rho}^{(n)}_{\mathrm{R}}}\rangle = \Lambda_{n} \ket{{\rho}^{(n)}_{\mathrm{R}}}\rangle, 
\end{align}
for $n=1, \cdots, \mathrm{dim}~\mathcal{L}$. 
As demonstrated in Sec.~\ref{sec:result}, eigenmodes and eigenvalues of the Liouvillian exhibit a strong dependence on boundary conditions.

\subsection{Two-body loss process}
\begin{figure}[t]
    \centering
    \includegraphics[width=1\hsize,clip]{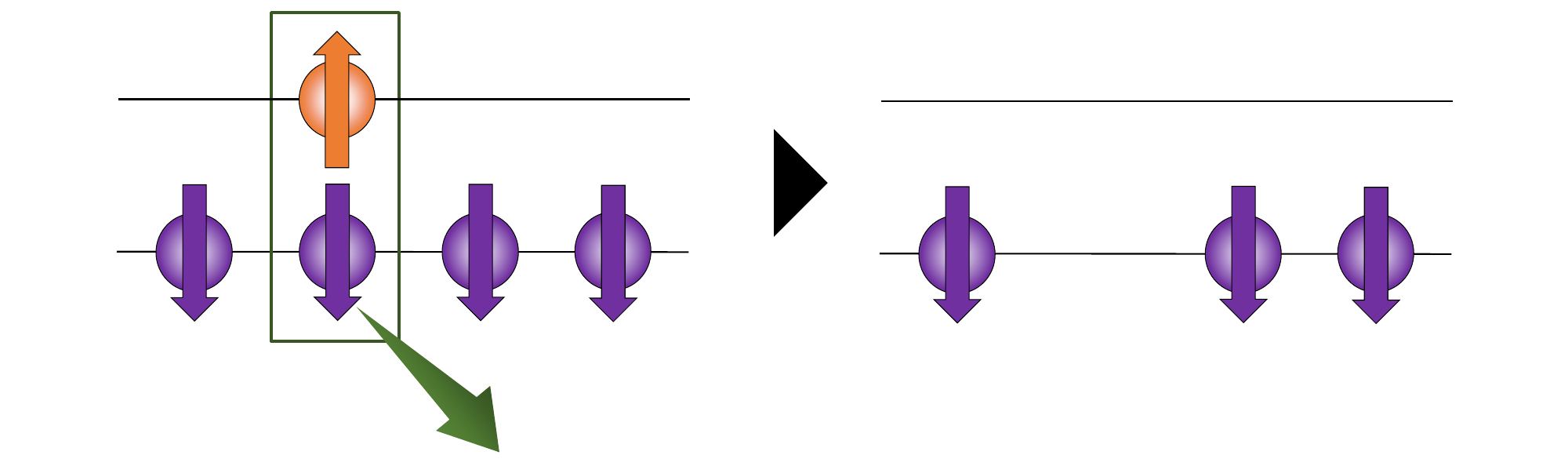}
    \caption{\rev{Schematic illustration of the two-body loss process for an initial state with $N_{\uparrow}=1$.
Due to the jump operator of the two-body loss $\mathcal{L}_{\mathrm{J}}$ given in Eq.~(\ref{eq:jump}), a fermion with an up-spin and that with a down-spin form pairs and are scattered into environments.}
        Down-spin configurations are changed from $\{n_{\downarrow}\}=\{1,1,1,1\}$
        to $\{n_{\downarrow}\}'=\{1,0,1,1\}$ after the two-body loss process.
    }
    \label{fig:situation}
\end{figure}
Let us consider the two-body loss process as illustrated in Fig.~\ref{fig:situation}.
\rev{\res{First,} we consider an initial state where only one fermion is in an up-spin state with down-spin configurations $\{ n_\downarrow \}$.}
\res{
As the commutation relation}
\begin{align}
    [\mathcal{L}_{0},N_{\uparrow}\otimes {N_{\uparrow}}]
    =  [\mathcal{L}_{0},n_{jb\downarrow}\otimes{n}_{jb\downarrow}]
    =0
\end{align}
indicates that the density matrix $\ket{\rho}\rangle$ is labeled by the total number of fermions in up-spin states $N_{\uparrow}$ and down-spin configurations $\{n_{\downarrow} \}$, the density matrix for $N_{\uparrow}=1$ with down-spin configurations $\{ n_{\downarrow} \}$ is spanned by the basis
\begin{align}
\label{eq:N=1}
     \ket{(N_{\uparrow}=1)}\rangle  = 
    c^{\dagger}_{j_{1}\alpha\uparrow}
    \otimes
    {c}^{\dagger}_{j_{2}\beta\uparrow}~
    \ket{\{n_{\downarrow}\}}\otimes \ket{{\{{n}_{\downarrow}\}}}
\end{align}
for $j_{1},j_{2}=1, \cdots, L $,\hspace{1mm} $\alpha,\beta= a,b$
in the absence of the jump operator $\mathcal{L}_{\mathrm{J}}$.
\res{Then, owing to the jump operator $\mathcal{L}_{\mathrm{J}}$ that describes the two-body loss process, a fermion in an up-spin state and that in a down-spin state form pairs and are scattered out into environments.}
For example, when the total number of down spins in the initial state is $N_{\downarrow}=4$,
the two-body loss process changes down-spin configurations
into one of the following four (see Fig.~\ref{fig:situation})
\begin{align}
    &\{n_{\downarrow}\}=\{1,1,1,1\} \nonumber \\
    \rightarrow~
    &\{n_{\downarrow}\}'=\{0,1,1,1\},\{1,0,1,1\},\{1,1,0,1\},\{1,1,1,0\}.
\end{align}
Here, $\{n_{\downarrow}\}'$ denotes the down-spin configurations after the two-body loss process.
Then, the density matrix for $N_{\uparrow}=0$ state is spanned by the basis
\begin{align}
\label{eq:N=0}
    \ket{(N_{\uparrow}=0)}\rangle = 
    \ket{\{n_{\downarrow}\}'}\otimes \ket{{\{{n}_{\downarrow}\}'}}.
\end{align}
We construct the basis set $\{\ket{i}\otimes\ket{j}\}$, which spans the doubled Hilbert space $\mathcal{H}\otimes\mathcal{H}$ given in Eq.~\eqref{eq:map},
by identifying the basis set $\{\ket{i}\otimes\ket{j}\}$ with 
the basis sets combining $\{ \ket{(N_{\uparrow}=1)}\rangle \}$ and $\{ \ket{(N_{\uparrow}=0)}\rangle\}$: 
\begin{align}
\label{eq:basis-set}
    \Bigl\{ \ket{i}\otimes\ket{j} \Bigr\}=\Bigl\{\ket{(N_{\uparrow}=1)}\rangle,
    \ket{(N_{\uparrow}=0)}\rangle \Bigr\}.
\end{align}
The matrix representation of the Liouvillian with respect to these bases
$\{\ket{i}\otimes\ket{j}\}$ takes the form
\begin{align}
    \mathcal{L} =
    \left(
    \begin{array}{c|c}
            \mathcal{L}_{0}^{(N_{\uparrow}=1)}   &    \\ \hline
            \mathcal{L}_{J}    & \mathcal{L}_{0}^{(N_{\uparrow}=0)} \\ 
        \end{array}
    \right).
    \label{eq:blocktri}
\end{align}
Here, $\mathcal{L}_{0}^{(N_{\uparrow}=1)}$ [$\mathcal{L}_{0}^{(N_{\uparrow}=0)}$] denotes the matrix representation
of $\mathcal{L}_{0}$ for $N_{\uparrow}=1$ [$N_{\uparrow}=0$] sector.
It is known that the block triangular structure is a general property of the Liouvillian for particle losses~\cite{Torres-PRA-2014, nakagawa-PRL-2021}.
This structure simplifies the calculation of the winding number of the Liouvillian \res{as shown in Sec.~\ref{sec:result}}.

\section{
  Topological invariant and skin mode of the Liouvillian
 }
\label{sec:definition}
\subsection{Topological invariant}
In this subsection, we first present the topological number defined by the Liouvillian superoperator, and then discuss the relation between the topological number and the Liouvillian skin effect.

First, we introduce the following topological invariant by using the Liouvillian superoperator
\begin{align}
    \nu(\Lambda_{\rm{ref}})= \oint_{0}^{2\pi}  \frac{d\theta }{2\pi i}
    \frac{d}{d\theta} \log\det
    \bigl[\mathcal{L}(\theta)-\Lambda_{\rm{ref}}\bigr], \label{eq:winding}
\end{align}
where we have imposed the twisted boundary condition only on the ket space.
Here, $\Lambda_{\mathrm{ref}}\in  \mathbb{C}$ denotes the reference point, \refirst{specifying the point-gap which we focus on.
The reference point $\Lambda_{\mathrm{ref}}$ is chosen from case by case in the same manner as the topology of the non-Hermitian Hamiltonian~\cite{Okuma-Topo, Kawabata-PRX}. $\mathcal{L}(\theta)$ is defined as}
\begin{align}
    \label{eq:twistHamiltonian}
    i \mathcal{L}(\theta)
    = {H}_{\rm{eff}} (\theta)\otimes {I}
    -I \otimes {{H}}_{\rm{eff}}^{*}
    +
    i \sum\limits_{j} {L}_{j}(\theta)\otimes {{L}}_{j}^{*},
\end{align}
\res{
where operators ${H}_{\rm{eff}} (\theta)$ and ${L}_{j}(\theta)$ are defined by multiplying $e^{\pm i\theta}$ to the hopping term at the boundary, e.g., 
$c^\dagger_{1\alpha\sigma} c_{L\alpha'\sigma'} $ is replaced by $c^\dagger_{1\alpha\sigma} c_{L\alpha'\sigma'}e^{i\theta}$.
More precisely, in the case of the Falicov-Kimball model with the two-body loss given in Eq.~\eqref{eq:eff-Hamil}, $H_{\mathrm{eff}}(\theta)$ is written by
\begin{align}
    H_{\mathrm{eff}}(\theta)=H_{\mathrm{eff}}^{\mathrm{bulk}} 
    + H_{\mathrm{eff}}^{\mathrm{edge}}(\theta)
\end{align}
where $H_{\mathrm{eff}}^{\mathrm{bulk}}$ is the Hamiltonian in the bulk, which is independent of $\theta$ and is written down as
\begin{align}
    H_{\mathrm{eff}}^{\mathrm{bulk}} = \sum\limits_{\langle ij \rangle^{'} \alpha \beta} h_{ i\alpha j\beta}
    {c}^{\dagger}_{i\alpha\uparrow}{c}_{j\beta\uparrow}
    + (U-i\gamma) \sum\limits_{j=1}^{L} {n}_{j b \uparrow} {n}_{j b \downarrow}.
\end{align}
The summation ${\langle ij \rangle^{'}}$  runs over all pairs of nearest neighbor sites $i$ and $j$,
excluding the hopping at the boundary between site $1$ and site $L$.
The boundary term of the Hamiltonian $H_{\mathrm{eff}}^{\mathrm{edge}}(\theta)$ is given by
\begin{align}
    H_{\mathrm{eff}}^{\mathrm{edge}}(\theta) = \sum\limits_{\alpha\beta}
    (h_{1\alpha L\beta} c_{1\alpha\uparrow}^{\dagger}c_{L\beta\uparrow}e^{i\theta} + \mathrm{h.c.}).
\end{align}
Here, $h_{1\alpha L\beta}$ is the hopping Hamiltonian between site $1$ and site $L$.
Since we consider the on-site dissipator given in Eq.~(\ref{eq:dissipator}), the Lindblad operator is independent on $\theta$, i.e. $L_{j}(\theta)=L_{j}$.
Because of the relation $\mathcal{L}(\theta)=\mathcal{L}(\theta+2\pi)$, the winding
number $\nu(\Lambda_{\mathrm{ref}})$ given in Eq.~(\ref{eq:winding}) is quantized. Hereafter, when the winding number given in Eq.~\eqref{eq:winding} takes a nonzero value, we denote that the point-gap topology of the Liouvillian is nontrivial.
}

Second, we discuss the relation between the topological number defined in Eq.~\eqref{eq:winding} and the Liouvillian skin effect.
Even in the single-particle system, the topological characterization of the Liouvillian skin effect has not been accomplished so far.
Importantly, in the single-particle system, the topological invariant $\nu(\Lambda_{\mathrm{ref}})$ defined in Eq.~(\ref{eq:winding}) gives the characterization of the Liouvillian skin effect, provided that the Lindblad operator is given by the asymmetric hopping (see Appendix~\ref{app:Haga} for details).
In Appendix~\ref{app:Haga}, we compute the topological invariant $\nu(\Lambda_{\mathrm{ref}})$ defined in Eq.~(\ref{eq:winding}) analytically and discuss the validity of the characterization of the Liouvillian skin effect in the single-particle system.
Significantly, the topological invariant $\nu(\Lambda_{\mathrm{ref}})$ can be computed even in many-body systems.
\refirst{Such a definition of topological invariant, which is independent of momentum, has already been introduced for the non-Hermitian skin effect in many-body systems~\cite{Kawabata-Manybody-PRB-2022, Zhang-Song-PRB-2022, Faugno-PRL-2022}.}
In the following section, we numerically calculate the topological invariant $\nu(\Lambda_{\mathrm{ref}})$ and observe the nontrivial value of the topological invariant $\nu(\Lambda_{\mathrm{ref}})$ corresponding to the Liouvillian skin effect in many-body systems~\footnote{In the case of the non-Hermitian skin effect in many-body systems, the rigorous proof of the relation between nonzero topological number and the emergence of skin effect has not been clarified. Some research has shown this relation by analyzing the specific model.  Similarity, we demonstrate that our topological number takes nontrivial value corresponding to the Liouvillian skin effect in many-body systems by analyzing the Falicov-Kimball model in the main text.}.

\subsection{Skin mode of the Liouvillian}
\res{In this subsection, we first introduce the right-state particle density of the $n$-th eigenmode of the Liouvillian superoperator $\mathcal{L}$ as $\Delta_{j\alpha\sigma}^{(n)}$, which measures the degree of localization of eigenmodes of the Liouvillian superoperator in many-body systems. 
Then, we show that in the single-particle system, the right-state particle density reduces to the diagonal element of the right eigenmode, which is used as the characterization of the Liouvillian skin effect in single-particle systems in Ref.~\cite{Haga-PRL}.
Finally, we show that when the right eigenmode is written by the right eigenstate of the effective Hamiltonian, the right-state particle density gives the particle density, which is used as the characterization of the non-Hermitian skin effect in many-body systems in Refs.~\cite{Faugno-PRL-2022, Kawabata-Manybody-PRB-2022}.
}

\res{
First, we define the following right-state particle density of the $n$-th eigenmode of the Liouvillian superoperator $\mathcal{L}$ to quantify the degree of localization of the eigenmode of the Liouvillian in many-body systems:
\begin{align}
    \label{eq:skin-mode}
    \Delta _{l}^{(n)}
    =
    \langle\bra{J} c^{\dagger}_{l}c_{l}
    \otimes {I}
    \ket{\rho_{\mathrm{R}}^{(n)}}\rangle
    =
    \langle\bra{J}
    I \otimes
    {c}^{\dagger}_{l}{c}_{l} \ket{\rho_{\mathrm{R}}^{(n)}}\rangle
\end{align}
{with $l$ denoting the set of $j$,$\alpha$ and $\sigma$, i.e. $l=j\alpha\sigma$.}
Here, $\ket{{J}\rangle}$ is the identity operator defined by $\ket{{J}\rangle}=\sum_{j}\ket{j}\otimes\ket{{j}}$, and 
$ \ket{\rho_{\mathrm{R}}^{(n)}}\rangle$ is the $n$-th right eigenmode of the Liouvillian $\mathcal{L}$ that satisfies the eigenvalue equation given in Eq.~(\ref{eq:eigenvalue-Liouvillian}).
We note that the right-state particle density is not identical to the ordinary particle density, which is observable and takes real values.
Specifically, the right-state particle density is complex-valued, which is introduced to measure the degree of localization of eigenmodes in many-body systems.
}

Next, we show that the right-state particle density defined in Eq.~(\ref{eq:skin-mode}) reduces to the diagonal element of the right eigenmode in the single-particle system.
We note that the right-state particle density of the $n$-th eigenmode given in Eq.~(\ref{eq:skin-mode}) is expressed as
\begin{align}
\label{eq:density-mat}
    \Delta_{l}^{(n)} = \Tr[c^{\dagger}_{l}c_{l}{\rho_{\mathrm{R}}^{(n)}}],
\end{align}
where we have used the following relation
\begin{align}
    \langle{\bra{J}} A\otimes {I}  \ket{\rho_{\mathrm{R}}^{(n)}}\rangle
     & = \sum\limits_{jkl}
    \bra{{j}} \otimes \bra{j}~
    \Bigl(
    A\ket{k} \otimes \ket{{l}}
    \Bigr)
    \rho_{\mathrm{R},{kl}}^{(n)} \nonumber                                 \\
     & = \sum\limits_{jkl} \delta_{jl}~\bra{j}A\ket{k}
     \rho_{\mathrm{R},{kl}}^{(n)}
    \nonumber                                          \\
     & = \Tr[A \rho_{\mathrm{R}}^{(n)}].
\end{align}
\res{
Now, we show that the definition given in Eq.~(\ref{eq:skin-mode}) reduces to
the diagonal element of the right eigenmode of the Liouvillian in the single-particle system.
We take the single-particle basis $\ket{S}$, which is generated by acting the creation operator to the vacuum $\ket{\mathrm{vac}}$ as $\ket{S}=c^{\dagger}_{S}\ket{\mathrm{vac}}$ with $S$ denoting the set of $j$,$\alpha$ and $\sigma$, i.e. $S = j\alpha\sigma$.
Then the $n$-th right eigenstate of the density matrix  $\rho^{(n)}_{\mathrm{R}}$ is expanded by using the single-particle state $\ket{S}$ as
\begin{align}
\label{eq:density-mat-single}
    \rho^{(n)}_{\mathrm{R}} = \sum\limits_{ST} \rho_{\mathrm{R}, ST}^{(n)} \ket{S}\bra{T}.
\end{align}
Here, $\rho^{(n)}_{\mathrm{R},ST} \in \mathbb{C}$ is expansion coefficient.
By substituting Eq.~\eqref{eq:density-mat-single} into Eq.~\eqref{eq:density-mat},
we obtain 
\begin{align}
\label{eq:density-single}
    \Delta_{S}^{(n)} = \rho_{\mathrm{R}, SS}^{(n)}.
\end{align}}
Thus the right-state particle density defined in Eq.~\eqref{eq:density-mat} is a generalization of the diagonal element of the right eigenmode, which measures the degree of localization of the eigenmode of the Liouvillian in the single-particle system.

Finally, we show that when the right eigenmode $\rho_{\mathrm{R}}^{(n)}$ is written by the right eigenstate of the effective Hamiltonian $H_{\mathrm{eff}}$ as $\ket{\varphi_{\mathrm{R}}^{(n)}}$, 
the right-state particle density reduces to the particle density defined by
$n_{l} = \bra{\varphi_{\mathrm{R}}^{(n)}}c^{\dagger}_{l} c_{l} \ket{\varphi_{\mathrm{R}}^{(n)}}$, which is used as the characterization of the skin mode in non-Hermitian
many-body system in Refs~\cite{Kawabata-Manybody-PRB-2022,Faugno-PRL-2022}. 
We take the right and left eigenstate of the effective Hamiltonian as $\ket{\varphi_{\mathrm{R}}^{(n)}}$ and $\bra{\varphi_{\mathrm{L}}^{(n)}}$ that satisfy the eigenvalue equations
\begin{align}
    H_{\mathrm{eff}} \ket{\varphi_{\mathrm{R}}^{(n)}} = E_{n} \ket{\varphi_{\mathrm{R}}^{(n)}}   
\end{align}
and 
\begin{align}
     \bra{\varphi_{\mathrm{L}}^{(n)}} H_{\mathrm{eff}} = E_{n} \bra{\varphi_{\mathrm{L}}^{(n)}},    
\end{align}
respectively.
When we take the right eigenmode $\rho_{\mathrm{R}}^{(n)}$ as
\begin{align}
    \rho_{\mathrm{R}}^{(n)} = \ket{\varphi_{\mathrm{R}}^{(n)}}   \bra{\varphi_{\mathrm{R}}^{(n)}},    
\end{align}
the right-state particle density given in Eq.~\eqref{eq:density-mat} becomes
\begin{align}
\label{eq:density-manybody}
     \Delta_{l}^{(n)} &= \Tr[{\rho_{\mathrm{R}}^{(n)}}c^{\dagger}_{l}c_{l}] \nonumber \\
     &= \sum\limits_{m}  \langle{\varphi_{\mathrm{L}}^{(m)}}\ket{\varphi_{\mathrm{R}}^{(n)}}   \bra{\varphi_{\mathrm{R}}^{(n)}}c^{\dagger}_{l}c_{l} \ket{\varphi_{\mathrm{R}}^{(m)}} \nonumber \\
     &= \sum\limits_{m}  \delta_{mn}   \bra{\varphi_{\mathrm{R}}^{(n)}}c^{\dagger}_{l}c_{l} \ket{\varphi_{\mathrm{R}}^{(m)}} = n_{l},
\end{align}
where we have used the biorthogonal relation in the third equality~\cite{Gong-Adv}. Therefore the right-state particle density is the generalization of the particle density $n_{l}$, which measures the degree of localization of the eigenstate in non-Hermian many-body systems.
In the following, we demonstrate the right-state particle density  $\Delta_{l}^{(n)}$ exhibits localization near the edge.
From Eqs.~\eqref{eq:density-single} and \eqref{eq:density-manybody}, we consider that the
right-state particle density is the proper definition to measure the degree of localization of eigenmodes of the Liouvillian in many-body systems.

\section{
  NUMERICAL RESULTS
 }
\label{sec:result}
In this section, we demonstrate that interactions can induce the Liouvillian skin effect by analyzing the Falicov-Kimball model introduced in Sec.~\ref{sec:model}.
First, in the noninteracting case ($U-i\gamma=0$), 
we show that the Liouvillian skin effect is absent.
Then, it is demonstrated that the complex-valued interaction $U-i\gamma$ induces the Liouvillian skin effect.
In the following discussion, we set $t_{h}=1$ as an energy unit.

\subsection{Noninteracting case}
\rev{First, we see that the Liouvillian skin effect is not observed in the noninteracting case $(U=\gamma=0)$.
In this case, the Liouvillian given in Eq.~(\ref{eq:Liouvillian}) becomes
\begin{align}
    \mathcal{L}_{\mathrm{free}} = -i\left({H}\otimes {I}
    -I \otimes {{H}}^{T}
    \right).
\end{align}
Because $\mathcal{L}_{\mathrm{free}}$ is skew-Hermitian, i.e. $\mathcal{L}^\dagger_{\mathrm{free}}=-\mathcal{L}_{\mathrm{free}}$,
its eigenvalues are purely imaginary or zero.}
In other words, all eigenvalues of the Liouvillian lie on the imaginary axis regardless of boundary conditions.
As a result, \rev{the point-gap topology} of the Liouvillian always becomes trivial because the winding number always takes zero.
Correspondingly, the eigenvalues and the eigenmodes of the Liouvillian are not sensitive to the boundary conditions
(for more details, see Appendix \ref{app:no-dissipation}).
Therefore, the Liouvillian skin effect is absent for the noninteracting system.

\subsection{Interacting case}
\begin{figure}[t]
    \centering
    \includegraphics[width=1\hsize,clip]{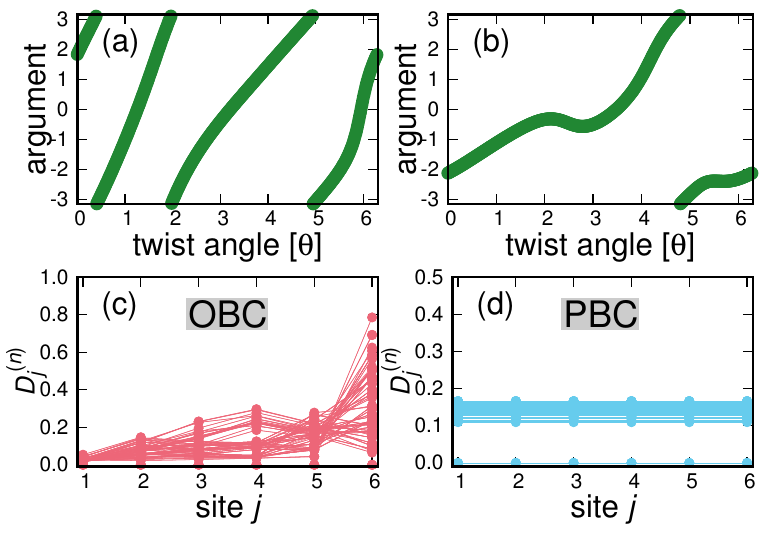}
    \caption{\rev{(a)[(b)] Argument of $\mathrm{det}[\mathcal{L}(\theta)-\Lambda_{\mathrm{ref}}]$ for $\Lambda_{\mathrm{ref}}=-0.5-0.8i$ [$\Lambda_{\mathrm{ref}}=-0.3-0.2i$]. Figure (a) [(b)] shows that the winding number takes $\nu=3$ [$\nu=1$].
    (c)[(d)] $D_{j}^{(n)}=\sum_{\alpha}|\Delta_{j\alpha\uparrow}^{(n)}|$ computed from the right-state particle density under OBC [PBC]. 
    The parameters are set to be $~L=6,~U=0.5,$ and $\gamma=1.0$.
     The configuration of fermions in the down-spin states is set to be $\{n_{\downarrow}\}=\{1,\cdots,1\}$. The density matrix is
    normalized as $\sum_{ij}|\rho_{\mathrm{R},ij}^{(n)}|^{2}=1$.
    }}
    \label{fig:winding}
\end{figure}

\rev{
Next, we demonstrate that the interaction $U-i\gamma$ makes the point-gap topology nontrivial and induces the Liouvillian skin effect.
Figures~\ref{fig:winding}(a) and (b) display the $\theta$ dependence of $\mathrm{det}[ \mathcal{L}(\theta)-\Lambda_{\mathrm{ref}}]$.
We see  that the winding number takes $\nu=3$ ($\nu=1$) for 
$\Lambda_{\mathrm{ref}}=-0.5-0.8i$ ($\Lambda_{\mathrm{ref}}=-0.3-0.2i$).
}
\rev{Now, we analyze the emergence of skin modes by comparing the results under OBC with those under periodic boundary conditions (PBC). 
Figure~\ref{fig:winding}(c) [(d)] displays 
$ D_j^{(n)}=\sum_{\alpha} |\Delta_{j\alpha\uparrow}^{(n)}|$ 
for OBC [PBC]. We note that the right-state particle density of the $n$-th eigenmode $\Delta_{j\alpha\uparrow}^{(n)}$ defined in Eq.~(\ref{eq:skin-mode}) takes a complex value.
Figure~\ref{fig:winding}(c) indicates that the eigenmodes are localized at the right edge under OBC. In contrast, such a localization cannot be observed under PBC.
These results demonstrate the emergence of skin modes of the Liouvillian.
We also note that the sensitivity of the eigenvalues to boundary conditions is also observed although it is smeared for small $L$ (for more details, see Appendix \ref{app:dep}).}
With the above results (see Fig.~\ref{fig:winding}), we conclude that interactions induce the Liouvillian skin effect though the system is subject to homogeneous two-body losses.
\refirst{
It should be noted that the Liouvillian skin effect occurs for $U=0$ and $\gamma \neq 0$. This skin effect is regarded as an interaction-induced skin effect since the nonzero $\gamma$ results in a two-body interaction [see Eq.~\eqref{eq:eff-Hamil}].}

\begin{figure}[t]
    \centering
    \includegraphics[width=1 \hsize,clip]{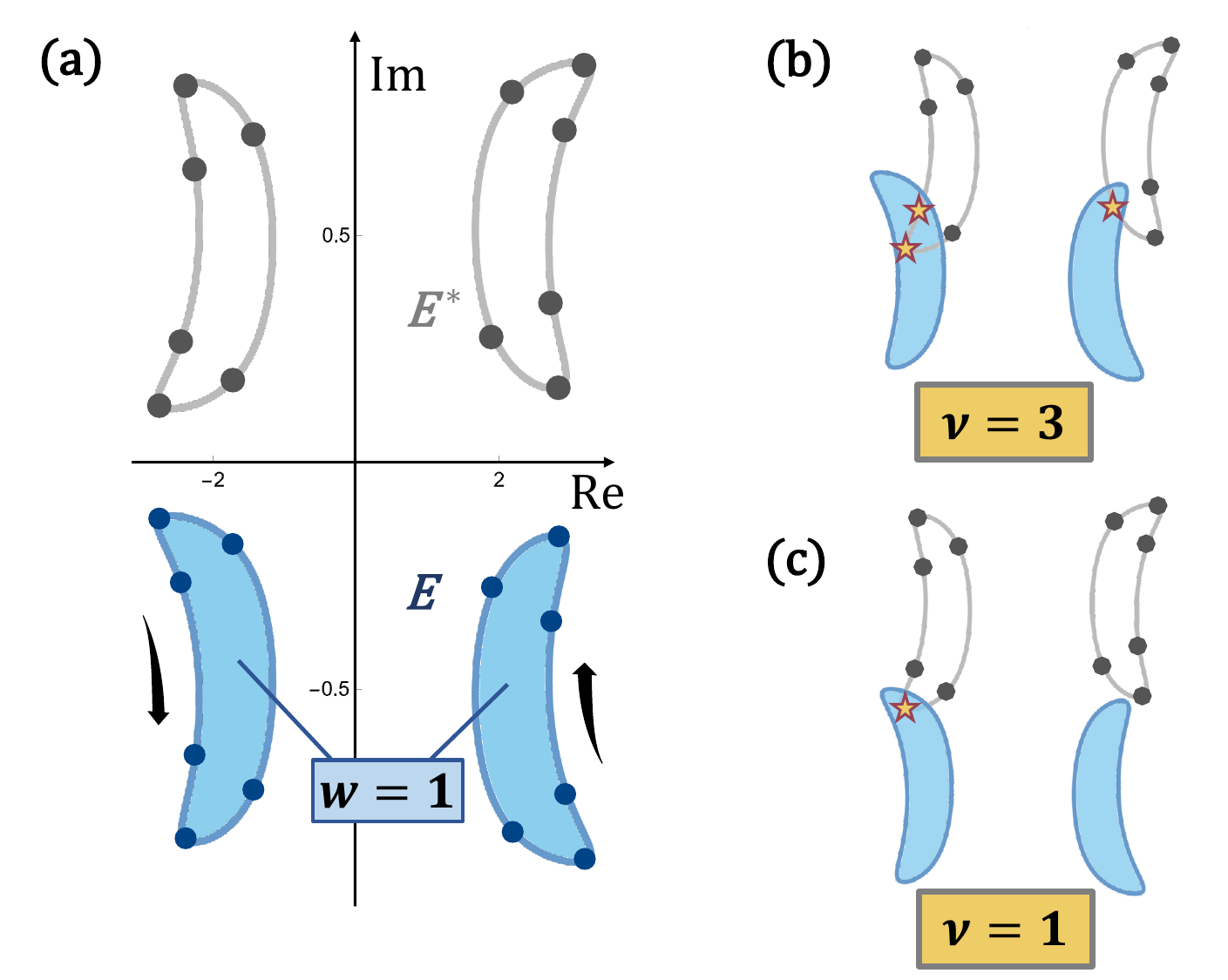}
    \caption{\rev{Schematic figure that describes the relationship between the winding number $\nu(\Lambda_{\mathrm{ref}})$ and $w(E_{\mathrm{ref}})$ given in Eq.~\eqref{eq:winding-relation}. 
        (a) Eigenvalues of the effective Hamiltonian $H_{\rm{eff}}$
        are indicated by blue dots.
        $E^{*}$ is the complex conjugation of the eigenvalue $E$ (gray dots).
        When $E_{\rm{ref}}$ is located inside the blue region, the winding number of the Hamiltonian equals one, i.e. $w(E_{\mathrm{ref}})=1$.
        (b),(c) Schematic figure
        of the origin of the nontrivial winding number.
        The winding number $\nu(\Lambda_{\mathrm{ref}})$ equals the number of 
        dots in the blue region indicated by the yellow star.
        The number of the yellow stars in panel (b) [(c)] corresponds to the winding number $\nu(\Lambda_{\mathrm{ref}})$ in Fig.~\ref{fig:winding}(a)[(b)].
        The constant shift $i\Lambda_{\rm{ref}}$ in Eq. \eqref{eq:winding-relation} is set to be $i\Lambda_{\rm{ref}} = 0.8-0.5i$ and $i\Lambda_{\rm{ref}} = ~0.2-0.3i$ for panel(b) and (c), respectively.
        The parameters are set to be $L=6,~U=0.5,~\gamma=1.0$. The configuration of fermions in the down-spin states is set to be $\{n_{\downarrow}\}=\{1,\cdots,1\}$. 
        }
    }
    \label{fig:relation}
\end{figure}

\rev{
Here, we comment on the relation between the winding number of the Liouvillian and that of the effective non-Hermitian Hamiltonian. As derived in Appendix~\ref{app:relationproof}, we obtain the following relation for the winding number defined in Eq.~\eqref{eq:winding}
\begin{align}
    \label{eq:winding-relation}
    \nu(\Lambda_{\rm{ref}}) = \sum\limits_{j} w(E_{\rm{ref}}=E_{j}^{*}+i \Lambda_{\rm{ref}}),
\end{align}
where $w(E_{\rm{ref}})$ is the winding number of the non-Hermitian Hamiltonian
\begin{align}
    \label{eq:wind-Hamiltonian}
    w(E_{\rm{ref}}) = \oint_{0}^{2\pi}\frac{d\theta}{2\pi i}
    \frac{d}{d\theta}\log \det [H_{\rm{eff}}(\theta)-E_{\rm{ref}}].
\end{align}
}\rev{
Here, $E_j$ denotes an eigenvalue of the non-Hermitian Hamiltonian $H_{\mathrm{eff}}$ defined in Eq.~(\ref{eq:eff-Hamil}).
Equation~(\ref{eq:winding-relation}) indicates that the winding number of the Liouvillian $\nu(\Lambda_{\rm{ref}})$ can be computed from the winding number of the effective non-Hermitian Hamiltonian $w(E_{\rm{ref}})$ with this model.
Here, we compute the winding number $\nu(\Lambda_{\rm{ref}})$ by making use of Eq.~(\ref{eq:winding-relation}).
First, we take the complex conjugate of the eigenvalue of the non-Hermitian Hamiltonian $E_j^{*}$ [see Fig.~\ref{fig:relation}(a)]. 
Then, we shift $E_j^{*}$ by $i\Lambda_{\mathrm{ref}}$ and obtain $E_{\mathrm{ref}}=E^*_j+i\Lambda_{\mathrm{ref}}$. Because Eq.~(\ref{eq:winding-relation}) indicates that the summation of $w$ for all possible $E_{\mathrm{ref}}=E^*_j+i\Lambda_{\mathrm{ref}}$ results in the winding number of the Liouvillian, we obtain $\nu(\Lambda_{\mathrm{ref}})=3$ [$\nu(\Lambda_{\mathrm{ref}})=1$] for $\Lambda_{\mathrm{ref}}=-0.5-0.8i$  ($\Lambda_{\mathrm{ref}}=-0.3-0.2i$) [see Figs.~\ref{fig:relation}(b) and (c)]. We note that the winding number of the effective Hamiltonian takes one when $E_{\mathrm{ref}}$ is in the shaded region in Fig.~\ref{fig:relation}(a).
These results of the winding number $\nu(\Lambda_{\mathrm{ref}})$ are consistent with the results obtained by the direct computation of $\nu(\Lambda_{\mathrm{ref}})$ [see Figs.~\ref{fig:winding}(a) and (b)].
}

\subsection{Dynamical properties}
\begin{figure}[t]
    \begin{center}
        \includegraphics[width=1\hsize,clip]{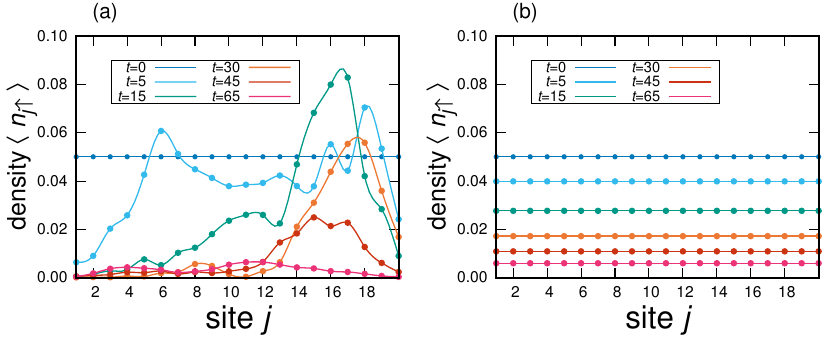}
        \caption{(a) [(b)] Time evolution of the particle density under OBC [PBC].
        The parameters are set to be $L=20, ~U=0.1,~ \gamma=0.1$. 
        The configuration of fermions in the down-spin states is set to be $\{n_{\downarrow}\}=\{1,\cdots,1\}$.
            The particle is uniformly distributed in orbital $a$ in the initial state. Only under OBC, the anomalous localization of the particle density is observed.
        }
        \label{fig:dynamics}
    \end{center}
\end{figure}

\rev{
In this subsection, we show that the Liouvillian skin effect significantly affects the dynamics of the particle density.} 
We assume the total number of particles in the up-spin states equals to one in the initial state, i.e., 
\begin{align}
    \bra{\Psi(t=0)} N_{\uparrow} \ket{\Psi(t=0)}=1,
\end{align}
where the wavefunction in the initial state $\ket{\Psi(t=0)}$ reads
\begin{align}
    \label{eq:uniform}
    \ket{\Psi(t=0)} =  \frac{1}{\sqrt{L}} \sum_{j=1}^{L}c^{\dagger}_{ja\uparrow}
    \ket{\{n_{\downarrow}\}}.
\end{align}
Here, we have assumed that the particle in orbital $a$ is uniformly distributed in the initial state.
The expectation value of the particle density in the up-spin state at time $t$ is given by
\begin{align}
    \label{eq:density}
    \langle n_{j\uparrow}(t) \rangle
    & =\sum_{\alpha} \Tr \left[{n}_{j\alpha\uparrow}{\rho}(t)\right]
      \nonumber    \\
    &=\sum_{\alpha}\langle\bra{J}c^{\dagger}_{j\alpha\uparrow}c_{j\alpha\uparrow}
    \otimes {I} \ket{\rho(t)}\rangle
     .
\end{align}
Then, the time evolution of the density matrix reads
\begin{align}
    \label{eq:dynamics}
    \ket{\rho(t)}\rangle= e^{\mathcal{L}t}\ket{\rho(t=0)}\rangle.
\end{align}
By using the wavefunction in the initial state $\ket{\Psi(t=0)}$,
the density matrix at time $t=0$ is given by 
\begin{align}
\label{eq:density-mat-ini}
    \ket{\rho(t=0)}\rangle = \ket{\Psi(t=0)} \otimes \ket{\Psi(t=0)}.
\end{align}
Here, $\ket{\Psi(t=0)} \otimes \ket{\Psi(t=0)}$ is defined by the following mapping
\begin{align}
    &\ket{\Psi(t=0)}\bra{\Psi(t=0)} = \sum\limits_{ij}\Psi_{ij}(t=0) \ket{i}\bra{j} \nonumber \\
    \mapsto
    &\ket{\Psi(t=0)} \otimes \ket{\Psi(t=0)} = \sum\limits_{ij}\Psi_{ij}(t=0) \ket{i}\otimes\ket{j}
\end{align}
where $\Psi_{ij}(t=0)$ is the matrix element of $\ket{\Psi(t=0)}\bra{\Psi(t=0)}$ and $\ket{i}\otimes\ket{j}$ is the element in the basis set given in Eq.~\eqref{eq:basis-set}.
Now, we numerically calculate the expectation value of the particle density given in Eq.~\eqref{eq:density} considering Eqs.~\eqref{eq:dynamics} and \eqref{eq:density-mat-ini}.
\rev{
We note that, since the initial state only has one fermion in an up-spin state, the dynamics is computed only from $H_{\mathrm{eff}}$~\cite{Yoshida-Hironobu-2023}.
Figure~\ref{fig:dynamics} displays the time dependence of the expectation value $\langle n_{j\uparrow}(t)\rangle$.
Under OBC, wee see that the particle is accumulated near the right boundary as shown in Fig.~\ref{fig:dynamics} (a). In contrast, under PBC, we find that the particle density decreases uniformly due
to the dissipation as shown in Fig.~\ref{fig:dynamics} (b).
}

\rev{
The above significant dependence of $\langle n_{j\uparrow} \rangle $ on boundary conditions can be understood in terms of the right-state particle density of the $n$-th eigenmode of the Liouvillian $\Delta_{j\alpha\sigma}^{(n)}$.
} First, we expand the initial density matrix
$\ket{{\rho}(0)}\rangle=\sum_{n}a_{n}\ket{\rho^{(n)}_{\mathrm{R}}}\rangle$ by using eigenmode of the Liouvillian. Then, by combining the eigenvalue equation of the Liouvillian given in Eq.~(\ref{eq:eigenvalue-Liouvillian}) and the time evolution of the density matrix given in Eq.~(\ref{eq:dynamics}), we obtain the particle density at time $t$ as
\begin{align}
    \label{eq:particle-eigenmode}
    \langle n_{j\uparrow} (t) \rangle =
    \sum\limits_{n, \alpha}
    e^{\Lambda_{n}t}a_{n}
    \Delta_{j\alpha\uparrow}^{(n)}.
\end{align}
Thus the particle accumulation quantitatively originates from the anomalous localization of $\Delta_{j\alpha\uparrow}^{(n)}$ (see Fig.~\ref{fig:winding}).
\refirst{Moreover, in the presence of the Liouvillian skin effect, we find the two-step relaxation process (see Appendix~\ref{app:two-step}).}

\refirst{We note that the particle accumulation due to the Liouvillian skin effect is different from the 
topological pumping~\cite{Lohse-Nature-2016, Nakajima-Nature-2016}.
In contrast to the topological pumping observed under the twisted boundary condition, the charge accumulation due to the skin modes is observed under the OBC. We twist the boundary condition only for computation of the topological invariant [Eq.~\eqref{eq:winding}].}

\section{
CONCLUSIONS
}
\label{sec:discussions}
\refirst{
In this paper, by introducing two-body loss into the one-dimensional correlated system, we have demonstrated that complex-valued interactions induce the Liouvillian skin effect.}
Specifically, by introducing the winding number constructed by the Liouvillian superoperator, we have elucidated that interactions make the point-gap topology nontrivial. 
Moreover, we have seen that eigenvalues and eigenmodes of the Liouvillian exhibit extreme sensitivity to boundary conditions. 
As a result, we have observed the particle accumulation around the right edge in transient dynamics only under OBC which is attributed to the emergence of the skin mode.

\rev{
As two-body losses have already been introduced in ytterbium atoms by using photoassociation techniques ~\cite{Tomita2, Honda}, 
our results can be tested in ultracold atoms.
The method to realize the Falikov-Kimball model is provided in Ref.~\cite{Semmler}, by introducing two species of atoms like $^{40}$K and $^{6}$Li, where mobile and immobile atoms are coupled via an on-site interaction.
We expect that the interaction-induced Liouvillian skin effect in our model can be observed in ultracold atoms.
}

\rev{
Recently, classifications of the Liouvillian superoperator have been actively conducted~\cite{Lieu-PRL-2020,Kawabata-Liouvillian-symmetry-2022}.
When the Hamiltonian preserves the inversion symmetry, 
the winding number defined in Eq.~(\ref{eq:winding}), which characterizes the nontrivial topological phase of the Liouvillian, can be trivial. 
Last but not least, it deserves to study the detailed relations 
between the symmetry and the topological number in a different Liouvillian, 
but we leave it for future work.
}


\section*{
Acknowledgments
}
\rev{
The authors are grateful to Naomichi Hatano, Hosho Katsura, Shuta Nakajima, Hironobu Yoshida, and Shin Kaneshiro for valuable discussions.
S.H. particularly acknowledges Masaya Nakagawa for fruitful discussions.
S.H. was supported by WISE Program, MEXT. K.Y. was supported by JSPS KAKENHI Grant-in-Aid for JSPS fellows Grant No.~JP20J21318 and JSPS KAKENHI Grants No.~JP23K19031.
K.Y. acknowledges the support by the research grant by Yamaguchi Educational and Scholarship Foundation.
This work was supported by JSPS KAKENHI Grants No.~JP22H05247 and No.~JP21K13850.
}

\appendix

\section{Symmetry constraint on the winding number}
\label{app:trivial-winding}
\rev{In this appendix, we discuss the relation between the symmetry of the 
Liouvillian and the winding number $\nu(\Lambda_{\mathrm{ref}})$. As we will see below, breaking the inversion symmetry of the Hamiltonian is essential for the existence of the nonzero topological number.
}
The winding number $\nu(\Lambda_{\mathrm{ref}})$ given in Eq.~(\ref{eq:winding}) becomes trivial when the Liouvillian
superoperator satisfies
\begin{align}
    \label{eq:inv-Liouvillian}
    \mathcal{U}\mathcal{L}(-\theta)\mathcal{U}^{\dagger} = \mathcal{L}(\theta).
\end{align}
Here, $\mathcal{U}$ is the unitary operator $(\mathcal{U}\mathcal{U}^{\dagger}=\mathcal{U}^{\dagger}\mathcal{U}=1)$.
We note that Eq.~(\ref{eq:inv-Liouvillian}) leads to $\nu(\Lambda_{\rm{ref}})=-\nu(\Lambda_{\rm{ref}})$.
This relation means that the point-gap topology of the Liouvillian is trivial $\nu(\Lambda_{\mathrm{ref}})=0$.

In the case of particle loss, this triviality $[\nu(\Lambda_{\mathrm{ref}})=0]$ originates from the symmetry of the Hamiltonian. Since the eigenvalue of the Liouvillian
is determined only by the effective Hamiltonian
$H_{\rm{eff}}$~\cite{Torres-PRA-2014}, the winding number given in Eq.~(\ref{eq:winding}) reduces to
\begin{align}
    \label{eq:winding_L0}
    \nu(\Lambda_{\rm{ref}}) = \oint_{0}^{2\pi} \frac{d\theta}{2\pi i}
    \frac{d}{d\theta} \log\det \bigl[\mathcal{L}_{0}(\theta)-\Lambda_{\rm{ref}}\bigr].
\end{align}
If the effective Hamiltonian ${H}_{\rm{eff}}$ satisfies the following relation
\begin{align}
    \label{eq:Heff-inv}
    {U}{H}_{\rm{eff}}(-\theta){U}^{\dagger} = {H}_{\rm{eff}}(\theta),
\end{align}
we can construct the unitary operator $\mathcal{U}$ as
\begin{align}
    \mathcal{U}={U}\otimes {V},
\end{align}
where ${U}{U}^{\dagger}={U}^{\dagger}{U}=1$ and  we have defined ${V}= {U}^{*}$.
Due to the relation $\mathcal{U}\mathcal{L}_{0}(-\theta)\mathcal{U}^{\dagger} = \mathcal{L}_{0}(\theta)$, we find that 
the winding number Eq.~(\ref{eq:winding_L0}) becomes zero.

Now, we discuss whether the Liouvillian given in Eq.~(\ref{eq:Liouvillian}) satisfies the condition given in Eq.~(\ref{eq:inv-Liouvillian}). \rev{The effective Hamiltonian
under twisted boundary conditions in real space is $H_{\mathrm{eff}}(\theta) = H_{\mathrm{eff}}^{\mathrm{bulk}} + H_{\mathrm{eff}}^{\mathrm{edge}}(\theta)$ where
$H_{\mathrm{eff}}^{\mathrm{bulk}}$ is written as
\begin{align}
    H_{\mathrm{eff}}^{\mathrm{bulk}} = H_{1} + H_{2} +H_{3} + H_{4}.
    \label{eq:twist-Hamiltonian-real-space}
\end{align}
Here, we have defined
\begin{align}
    &H_{1} = -2i t_{h} \sum\limits_{j=1}^{L}({c}^{\dagger}_{ja\uparrow}{c}_{jb\uparrow}-\mathrm{h.c.}), \nonumber \\
    &H_{2} = -0.25 t_{h} \sum\limits_{j=1}^{L-1}({c}^{\dagger}_{j+1 b\uparrow}{c}_{ja\uparrow}-{c}^{\dagger}_{j+1a\uparrow}{c}_{jb\uparrow}+\mathrm{h.c.}), \nonumber \\
    &H_{3} = t_{h} \sum\limits_{j=1}^{L-1}({c}^{\dagger}_{j+1a\uparrow}{c}_{ja\uparrow}-{c}^{\dagger}_{j+1b\uparrow}{c}_{jb\uparrow}+\mathrm{h.c.}), \nonumber \\
    &H_{4} = (U-i\gamma) \sum\limits_{j=1}^{L} {n}_{j b \uparrow} {n}_{j b \downarrow},
\end{align}
and
\begin{align}
        H_{\mathrm{eff}}^{\mathrm{edge}}(\theta) =&-0.25 t_{h} (e^{i\theta}{c}^{\dagger}_{1 b\uparrow}{c}_{La\uparrow}-e^{i\theta}{c}^{\dagger}_{1a\uparrow}{c}_{Lb\uparrow}+\mathrm{h.c.}) \nonumber \\
        &+ t_{h} (e^{i\theta}{c}^{\dagger}_{1a\uparrow}{c}_{La\uparrow}-e^{i\theta}{c}^{\dagger}_{1b\uparrow}{c}_{Lb\uparrow}+\mathrm{h.c.}).
\end{align}
}\rev{
The first term $H_{1}$ of the bulk Hamiltonian $H_{\mathrm{eff}}^{\mathrm{bulk}}$ [i.e., the first term of the Bloch Hamiltonian in Eq.~(\ref{eq:b2}) denoted by $2t_{h}$] violates the condition Eq.~(\Ref{eq:inv-Liouvillian}). In the absence of the first term, the effective  Hamiltonian preserves the inversion symmetry
defined by
\begin{align}
    \label{eq:inversion-symm}
    {P}{H}_{\rm{eff}}(-\theta){P}^{\dagger} = {H}_{\rm{eff}}(\theta),
\end{align}
where the inversion operator $P$ acts on the annihilation operator as ${P}{c}_{ja\sigma}{P}^{\dagger} = {c}_{L-(j-1)a\sigma},{P}{c}_{jb\sigma}{P}^{\dagger} =- {c}_{L-(j-1)b\sigma} $
and satisfies ${P}^{\dagger}{P}={P}{P}^{\dagger}={\mathbf{1}}$.
\rev{Then, we see that the inversion symmetry given in} Eq.~(\ref{eq:inversion-symm}) is nothing but the condition of the triviality of the winding number given in
Eq. (\ref{eq:Heff-inv}). Therefore, the Liouvillian superoperator $\mathcal{L}(\theta)$ given in Eq.~(\ref{eq:twistHamiltonian}) satisfies Eq. (\ref{eq:inv-Liouvillian})
in the absence of $H_{1}$.
The presence of $H_{1}$ breaks the condition Eq.~(\ref{eq:inversion-symm}), which leads to the violation of the condition Eq.~(\ref{eq:inv-Liouvillian}).
Hence, the nonzero winding number originates from the property of the Liouvillian.
In particular, the Hamiltonian breaks the inversion symmetry.}

\section{Topological characterization of the Liouvillian skin effect reported in Ref.~\cite{Haga-PRL}}
\label{app:Haga}
\rev{In this appendix, we show that the winding number $\nu(\Lambda_{\mathrm{ref}})$ defined in Eq.~(\ref{eq:winding}) characterizes the Liouvillian skin effect reported in Ref.~\cite{Haga-PRL}, which implies the validity of employing $\nu(\Lambda_{\mathrm{ref}})$ for characterizing the interaction-induced Liouvillian skin effect.
}
We consider the bosonic systems and assume that the Lindblad operators are given by
\begin{align}
    L_{j,l} & = \sqrt{t_{l}}b_{j}^{\dagger}b_{j+1}, \nonumber \\
    L_{j,r} & = \sqrt{t_{r}}b_{j+1}^{\dagger}b_{j},
\end{align}
which describe the stochastic hopping to the nearest neighbor sites.
Following the discussion in Ref.~\cite{Haga-PRL}, 
we assume that the Hamiltonian of the systems is zero, i.e. $H=0$.
The Liouvillian superoperator becomes
\begin{widetext}
  \begin{align}
    \mathcal{L}^{H=0}
     &= 
     \sum\limits_{j,\alpha} \Biggl[L_{j,\alpha}\otimes L^{*}_{j,\alpha} 
     - \frac{1}{2}\left( 
     L_{j,\alpha}^{\dagger} L_{j,\alpha} \otimes I
     +
     I \otimes L_{j,\alpha}^{T} L_{j,\alpha}^{*}
     \right) \Biggr]
     \nonumber \\
     & = \sum\limits_{j=1}^{L} \Biggl[
    t_{r} b_{j+1}^{\dagger}b_{j} \otimes {b}_{j+1}^{\dagger}{b}_{j}
    +
    t_{l} b_{j}^{\dagger}b_{j+1} \otimes {b}_{j}^{\dagger}{b}_{j+1}
    -\frac{t_{r}+t_{l}}{2} \left( 
     b_{j}^{\dagger} b_{j} \otimes I
     +
     I \otimes b_{j}^{\dagger} b_{j}
     \right)
     \Biggr].
  \end{align}
\end{widetext}
When we impose twisted boundary conditions
only on the ket space, the Liouvillian superoperator is expressed as
\begin{align}
\label{eq:Liou_HN_theta}
    \mathcal{L}^{H=0} (\theta) 
     &= \mathcal{L}^{H=0}_{\rm{bulk}} + 
     t_{r} e^{i\theta} b_{1}^{\dagger}b_{L} \otimes {b}_{1}^{\dagger}{b}_{L}
    +
    t_{l} e^{-i\theta}b_{L}^{\dagger}b_{1} \otimes {b}_{L}^{\dagger}{b}_{1},
\end{align}
where we have introduced the bulk term of the Liouvillian $\mathcal{L}^{H=0}_{\rm{bulk}}$ as
\begin{align}
    \mathcal{L}^{H=0}_{\rm{bulk}} =  &\sum\limits_{j=1}^{L-1} \Biggl[
    t_{r} b_{j+1}^{\dagger}b_{j} \otimes {b}_{j+1}^{\dagger}{b}_{j}
    +
    t_{l} b_{j}^{\dagger}b_{j+1} \otimes {b}_{j}^{\dagger}{b}_{j+1}
    \Biggr] \nonumber \\
    - &\sum\limits_{j=1}^{L} \Biggl[
     \frac{t_{r}+t_{l}}{2} \left( 
     b_{j}^{\dagger} b_{j} \otimes I
     +
     I \otimes b_{j}^{\dagger} b_{j}
     \right)
     \Biggr]
     .
\end{align}
In the following discussion, we focus on the single-particle diagonal subspace spanned by the basis $\{\ket{i}\otimes\ket{{i}}\}_{i=1,\cdots,L}$.
The matrix representation of the Liouvillian with respect to this basis is given by
\begin{align}
    \mathcal{L}^{H=0}(\theta) = 
    \left(
    \begin{array}{c|c|c|c}
            -(t_{l}+t_{r})                 & t_{l}  &        & t_{r}e^{i\theta} \\ \hline
            t_{r}             & \ddots & \ddots &                  \\ \hline
                              & \ddots & \ddots & t_{l}            \\ \hline
            t_{l}e^{-i\theta} &        & t_{r}  & -(t_{l}+t_{r})
        \end{array}
    \right).
    \label{eq:HN-matrix}
\end{align}
In this subspace, the action of the Liouvillian $ \mathcal{L}^{H=0}(\theta)$ is identical to that of the following Hamiltonian in the single-particle system
\begin{align}
\label{eq:HN}
    {H}_{\rm{HN}}(\theta)  = 
    &-\sum\limits_{j=1}^{L}
    \left(t_{l}+t_{r}\right) {c}_{j}^{\dagger}{c}_{j} + 
    \sum\limits_{j=1}^{L-1}
    \left(t_{l}{c}_{j}^{\dagger}{c}_{j+1}
    +
    t_{r}{c}^{\dagger}_{j+1}{c}_{j}
    \right)
    \nonumber \\
    &+
    t_{l}{c}_{L}^{\dagger}{c}_{1}e^{-i\theta}
    +
    t_{r}{c}^{\dagger}_{1}{c}_{L}e^{i\theta}.
\end{align}
We note that the matrix representation of the Hamiltonian $H_{\mathrm{HN}}(\theta)$ with respect to the basis $\{\ket{i}\}_{i=1,\cdots,L}$ gives the Eq.~\eqref{eq:HN-matrix}.
The Hamiltonian given in Eq. \eqref{eq:HN} is nothing but the Hatano-Nelson model~\cite{Hatano-Nelson1,Hatano-Nelson2,Hatano-Nelson-3}
under twisted boundary conditions.

\rev{
Now, we calculate the winding number $\nu(\Lambda_{\mathrm{ref}})$ defined in Eq.~(\ref{eq:winding}) for the 
Liouvillian $\mathcal{L}^{H=0}(\theta)$ in this subspace.}
First, we recover the translational invariance of the Hamiltonian by using the gauge transformation
${c}_{j} \rightarrow {c}_{j}e^{-i\frac{j}{L}\theta}$ as
\begin{align}
\label{eq:HN_twist}
    {H}_{\rm{HN}}(\theta)=
    \sum\limits_{j=1}^{L}
    \Bigl[& -(t_{l}+t_{r}){c}_{j}^{\dagger}{c}_{j} \nonumber \\
    &+ 
    t_{l}e^{-i\frac{\theta}{L}}{c}_{j}^{\dagger}{c}_{j+1}+
    t_{r}e^{i\frac{\theta}{L}}{c}^{\dagger}_{j+1}{c}_{j}
    \Bigr].
\end{align}
Then, we diagonalize Eq.~(\ref{eq:HN_twist}) as
\begin{align}
    {H}_{\rm{HN}}(\theta)
    =\sum\limits_{k} h_{\rm{HN}}\left({k+\frac{\theta}{L}}\right){c}^{\dagger}_{k}{c}_{k}
\end{align}
with
\begin{align}
    h_{\rm{HN}} \left({k}\right) =
    t_{r}e^{ik}+t_{l}e^{-ik} - (t_{l}+t_{r}).
\end{align}
\rev{
In the translational invariant single-particle system,
many-body topological invariant of non-Hermitian systems reduces to the following topological invariant defined in the momentum space}
~\cite{Kawabata-Manybody-PRB-2022,wind_sign}
\begin{align}
    \label{eq:winding-free}
    W (\Lambda_{\rm{ref}}) =  \oint_{0}^{2\pi}  \frac{dk}{2\pi i}
    \frac{d}{dk} \log \det
    [h_{\rm{HN}}(k)-\Lambda_{\rm{ref}}].
\end{align}
Finally, in a similar way to Eq.~(\ref{eq:winding-free}), we can compute the winding number $\nu(\Lambda_{\mathrm{ref}})$ defined in Eq.~(\ref{eq:winding}) for $\mathcal{L}^{H=0}(\theta)$ given in Eq.~(\ref{eq:HN-matrix}) as
\begin{align}
\label{eq:wind-HN}
    \nu(\Lambda_{\mathrm{ref}}) = \sgn ~(t_{r}-t_{l}),
\end{align}
\rev{
where we set $\Lambda_{\rm{ref}}$ inside the region enclosed by the PBC spectrum.
}\rev{
As shown in Ref.~\cite{Haga-PRL}, the right-state particle density of the Liouvillian given in Eq.~\eqref{eq:skin-mode} exhibits the skin effect.
Therefore, the Liouvillian skin effect demonstrated in Ref.~\cite{Haga-PRL}
is characterized by the winding number defined in Eq.~(\ref{eq:winding}).
This fact supports the use of $\nu(\Lambda_{\mathrm{ref}})$ as a characterization of the interaction-induced Liouvillian skin effect as shown in the main text.
It should be noted that even for the presence $H$ in Ref.~\cite{Haga-PRL}, we numerically confirm that the winding number takes a nonzero value.}

\section{Absence of the Liouvillian skin effect in noninteracting systems}
\label{app:no-dissipation}
\begin{figure}[!h]
    \begin{center}
        \includegraphics[width=1\hsize,clip]{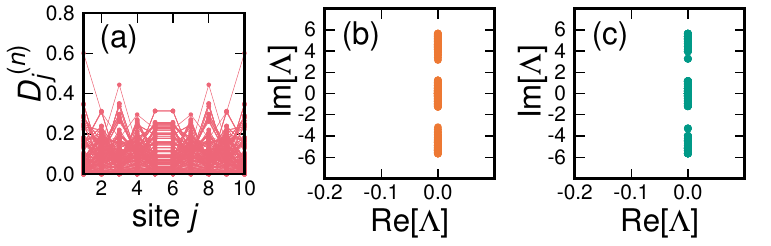}
    \end{center}
    \caption{(a) The right-state particle density of the Liouvillian under OBC.
        (b) [(c)] Eigenvalues of the Liouvillian under OBC [PBC] for the noninteracting case. The parameters are set to be $L=10,~ U=\gamma=0.0$.
        The configuration of fermions in the down-spin states is set to be $\{n_{\downarrow}\}=\{1,\cdots,1\}$.
    }
    \label{fig:nonint}
\end{figure}
Here we numerically show that the Liouvillian skin effect is absent when the systems do not have interactions ($U-i\gamma=0$). In this case, the time evolution of the density matrix is described by the von Neumann Equation.
Figure~\ref{fig:nonint}(a) shows that the eigenmodes of the Liouvillian do not exhibit the skin effect for the noninteracting case.
As mentioned in Sec.~\ref{sec:result}, all eigenvalues of the Liouvillian lie on the imaginary axis and are insensitive to boundary conditions [see Figs.~\ref{fig:nonint}(b) and \ref{fig:nonint}(c)].
Thus, the Liouvillian skin effect does not occur in the noninteracting system.
 
\section{\res{Sensitivity of eigenvalues of the Liouvillian to boundary conditions}}
\label{app:dep}
\begin{figure}[H]
    \centering
    \includegraphics[width=1\hsize,clip]{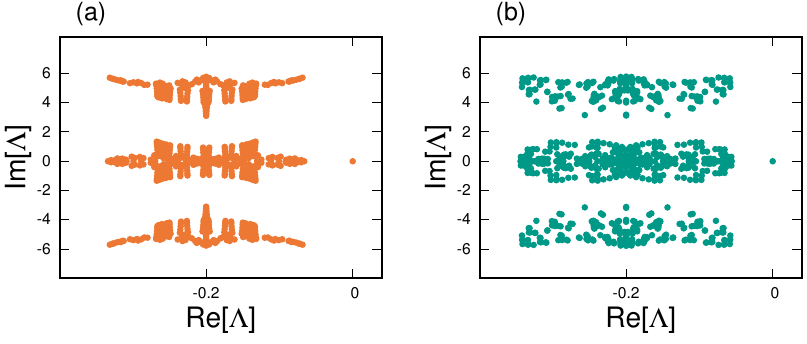}
    \caption{(a)[(b)] Eigenvalues of the Liouvillian under OBC [PBC]. The parameters are set to be $L=14$,~$U=0.1$, and $\gamma=0.2$.
    The configuration of fermions in the down-spin states is set to be $\{n_{\downarrow}\}=\{1,\cdots,1\}$.
    }
    \label{fig:eigenvalue-Liouvillian}
\end{figure}
\rev{
In this appendix, we show that eigenvalues of the Liouvillian exhibit the sensitivity to boundary conditions. 
Under OBC as shown in Fig.~\ref{fig:eigenvalue-Liouvillian}(a), the eigenvalues form a line-like structure, which is in contrast to the case of PBC shown in Fig.~\ref{fig:eigenvalue-Liouvillian}(b). 
Such sensitivity is a signal of the Liouvillian skin effect.
We note that since the steady state is $N_{\downarrow}$-fold degenerate regardless of boundary conditions, the eigenvalues corresponding to the steady state do not exhibit the Liouvillian skin effect.
}

\section{Derivation of the relation between winding numbers given in Eq.~(\ref{eq:winding-relation})}
\label{app:relationproof}
\rev{
In this appendix, we derive the relation between the winding number $\nu(\Lambda_{\mathrm{ref}})$ defined by the Liouvillian superoperator and $w(E_{\mathrm{ref}})$ defined by the Hamiltonian given in Eq.~(\ref{eq:winding-relation}).
First, we recall that in the case of particle losses, the Liouvillian takes the following block triangular structure
\begin{align}
    \mathcal{L} =
    \left(
    \begin{array}{c|c}
            \mathcal{L}_{0}^{(N_{\uparrow}=1)}(\theta)   &    \\ \hline
            \mathcal{L}_{\mathrm{J}}(\theta)  & \mathcal{L}_{0}^{(N_{\uparrow}=0)}(\theta) \\ 
        \end{array}
    \right).
    \label{eq:sector-blocktri}
\end{align}
We note that, since the Lindblad operator given in Eq.~\eqref{eq:dissipator} has no hopping term between site $1$ and site $L$, the jump term is independent on $\theta$, i.e. $\mathcal{L}_{\mathrm{J}}(\theta) =\mathcal{L}_{\mathrm{J}}$.
For a block triangular matrix, the following relation holds:}
\begin{align}
    \det
    \left(
    \begin{array}{c|c}
            A &   \\ \hline
            B & C \\
        \end{array}
    \right)
    =
    \det A \det C.
    \label{eq:abc-blocktri}
\end{align}
Since $\mathcal{L}^{(N_{\uparrow}=0)}(\theta)$ is independent of $\theta$, we obtain
\begin{align}
      & d_{\theta} \log\det [
        \mathcal{L} (\theta) - \Lambda_{\rm{ref}}
    ] \nonumber               \\
    =
      & d_{\theta}\log \det [
        \mathcal{L}^{(N_{\uparrow}=1)}_{0} (\theta)-\Lambda_{\rm{ref}}
    ] \nonumber               \\
    = & d_{\theta} \log \det
    \mathcal{M}(\theta)
\end{align}
for $\Lambda_{\mathrm{ref}} \neq 0$, where 
\begin{align}
    \mathcal{M}(\theta) = H_{\mathrm{eff}}(\theta)\otimes{I} - I\otimes{H}_{\mathrm{eff}}^{*}
    -i \Lambda_{\rm{ref}} I \otimes {I}.
\end{align}
Then, we introduce $\mathcal{N}(\theta)$ to diagonalize $\mathcal{M}(\theta)$ defined by
\begin{align}
    \mathcal{N}(\theta) = S(\theta)\otimes{T},
\end{align}
where ${T}=S^{*}(\theta=0)$, and operator $S(\theta)$ which diagonalizes $H_{\mathrm{eff}}(\theta)$ as
\begin{align}
    S^{-1}(\theta)H_{\mathrm{eff}}(\theta)S(\theta) = {\rm{diag}}
    \Bigl(
    E_{1}(\theta), \cdots, E_{2L}(\theta)
    \Bigr).
\end{align}
A straightforward calculation results in
\begin{align}
    &\log \det \Bigl[
        \mathcal{N}^{-1} \mathcal{M}(\theta)~\mathcal{N}
        \Bigr] \nonumber \\
    &= \sum\limits_{i=1}^{2L}
    \log\det\prod_{j=1}^{2L}
    \biggl[
        E_{j}(\theta)-
        (E_{i}^{*}+i \Lambda_{\rm{ref}})
        \biggr].
\end{align}
Finally, we obtain the relation between $\nu(\Lambda_{\mathrm{ref}})$ and $w(E_{\mathrm{ref}})$ as
\begin{align}
    & \oint_{0}^{2\pi}   \frac{d\theta }{2\pi i}
    \frac{d}{d\theta} \log\det
    [\mathcal{L}(\theta)-\Lambda_{\rm{ref}}I\otimes{I}~]
    \nonumber \\
    &= \sum\limits_{j} w(E_{\rm{ref}}=E_{j}^{*}+i \Lambda_{\rm{ref}}),
\end{align}
which is nothing but Eq.~(\ref{eq:winding-relation}) in the main text.

\section{\refirst{Slowing down of the relaxation process under OBC}}
\label{app:two-step}
\refirst{There have been several research which focus on the relaxation process in the presence of the Liouvillian skin effect~\cite{Haga-PRL, Wang-PRB-2023, Samuel-arxiv-2023}.
In this appendix, we show that the relaxation process slows down under OBC in a finite system. 
\begin{figure}[H]
\centering
    \includegraphics[width=1\hsize,clip]{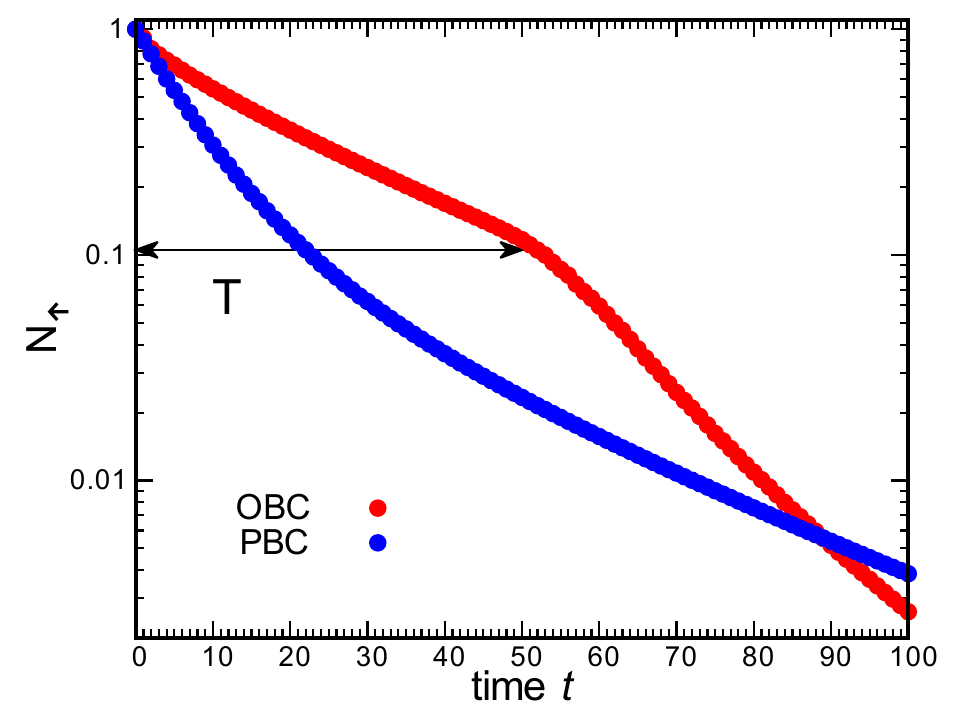}
\caption{Numerical result of the time dependence of the total number of the up-spin particles $N_{\uparrow}(t)$. 
$U=0.5, \gamma=0.1, L=60$. The configuration of fermions in the down-spin states is set to be $\{n_{\downarrow}\}=\{1,\cdots,1\}$. In the initial state, the particle is distributed at the left side of the system in the orbital $b$.
Times of the transition $T$ is denoted by the black arrow.}
\label{fig:two-step_relax}
\end{figure}
Figure~\ref{fig:two-step_relax} shows the time dependence of the total number of the particles in the up-spin state $N_{\uparrow}(t)$. 
By comparing the results under OBC and PBC, we find that the relaxation process is slowing down under OBC, which is originally discussed in Refs.~\cite{Haga-PRL}.
Under OBC, the relaxation speed changes after time $T$ in a finite system.
We calculate the system size dependence of the time of the transition time $T$.
\begin{figure}[!h]
\centering
    \includegraphics[width=1\hsize,clip]{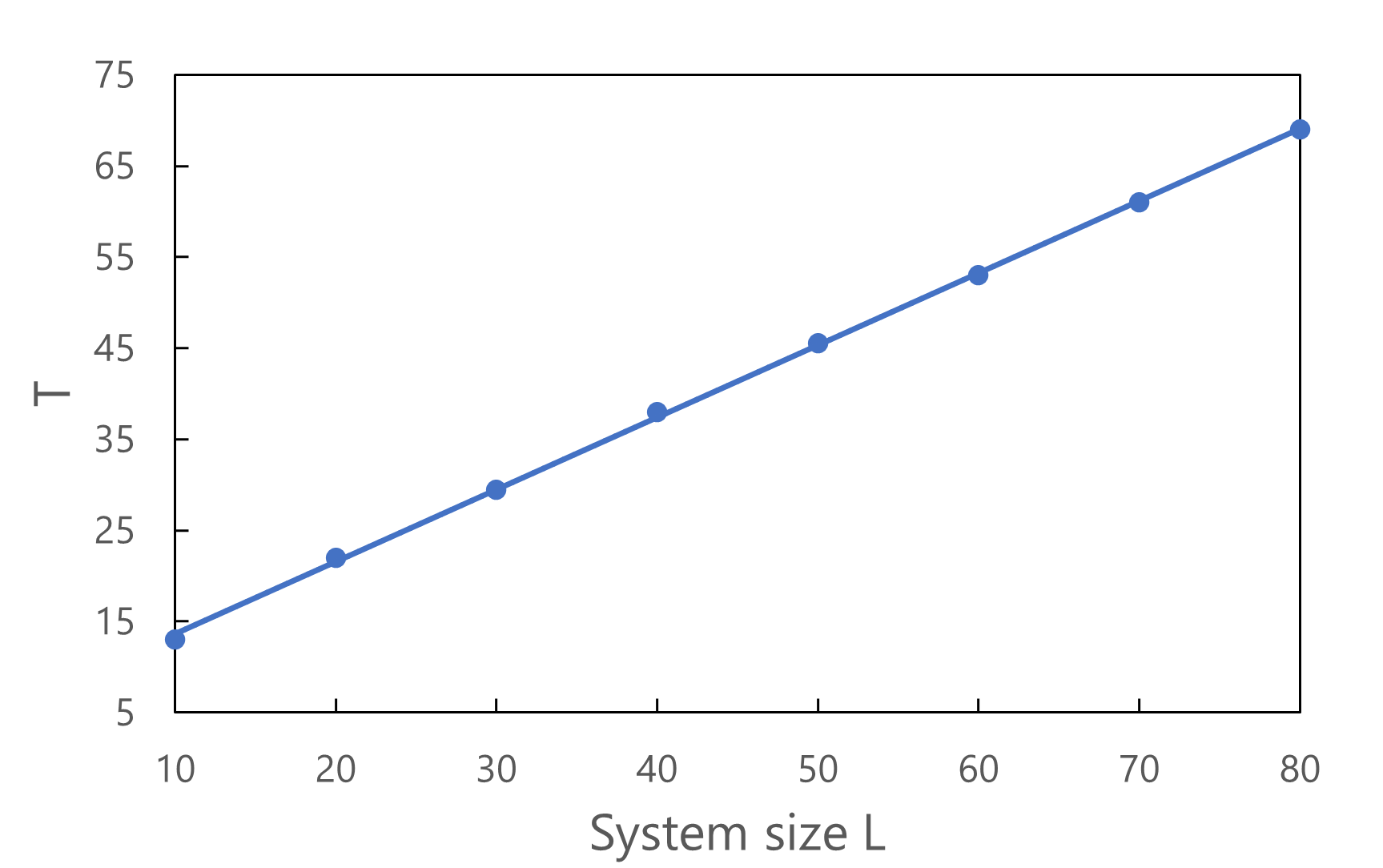}
\caption{The system size dependence of the transition time.
$U=0.5, \gamma=0.1$. In the initial state, the particle is distributed at the left side of the system in the orbital $b$.}
\label{fig:transition_size_dep}
\end{figure}
Figure~\ref{fig:transition_size_dep} indicates that the transition time $T$ and systems size are proportional$(T \propto L)$. This fact implies that slowing down the relaxation process occurs even when we increase the system size.
These results are consistent with the previous study of the dynamics with Liouvillian skin effect~\cite{Haga-PRL}.
}

\section{\res{Results for other configurations of down-spins in the initial state}}
\label{app:other-case}
\begin{figure}[!h]
    \centering
    \includegraphics[width=1\hsize,clip]{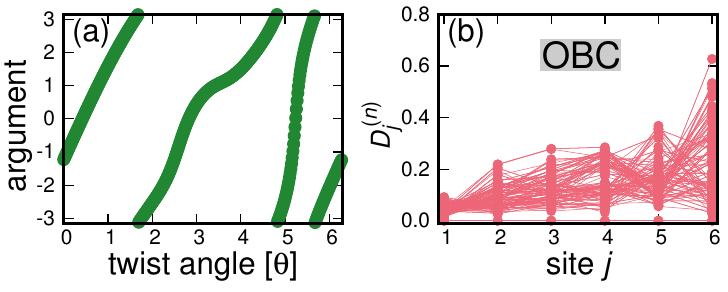}
    \caption{(a) Argument of $\det [\mathcal{L}(\theta)-\Lambda_{\rm{ref}}]$. 
    (b) The right-state particle density of the Liouvillian under OBC for other configuration cases of fermions in the down-spin states.
    The parameters are set to be $L=6$,~$U=0.3$, $\gamma=0.7$, and $\Lambda_{\rm{ref}}=-0.8-0.2i)$.
    The configuration of fermions in the down-spin states is set to be $\{n_{\downarrow}\}=\{1,1,1,1,0,1\}$.
    }
    \label{fig:eigenmode-defect}
\end{figure}
\res{
In the main text, the configuration of fermions in the 
down-spin states is set to be $\{n_{\downarrow}\}=\{1,\cdots,1\}$ in the initial state.
In this appendix, we show that the Liouvillian skin effect survives
for other configurations of down-spins in the initial state.
We set the configuration of fermions in the down-spin states to be $\{n_{\downarrow}\}=\{1,1,1,1,0,1\}$.
Then we numerically calculate the winding number given in Eq.~(\ref{eq:winding}).
Figure~\ref{fig:eigenmode-defect}(a) shows that the winding number
takes three.}
\rev{Moreover, eigenmodes of the Liouvillian exhibit the skin effect under OBC as shown in Fig.~\ref{fig:eigenmode-defect}(b).
We observe the dependence of eigenvalues on boundary conditions, which is similar to that presented in Appendix~\ref{app:dep}.
Therefore, the Liouvillian skin effect survives for other configurations of down-spins in the initial state.}

\section{\refirst{Continuous deformation of the effective Hamiltonian}}
\label{app:deform}
\refirst{
In this appendix, we perform the continuous deformation of the effective Hamiltonian, which supplies an intuitive understanding of why the eigenstate is localized at the right edge.
Non-Hermitian Hamiltonians $H_0 (k)$ and $H_1 (k)$ are defined to be topologically equivalent if and only if there exists a Hamiltonian satisfying and maintaining both relevant symmetry and the point-gap
$\det [H_\lambda(k)-E_{\mathrm{ref}}] \neq 0$ for  all $\lambda \in [0,1]$~\cite{Kawabata-PRX, Gong-PRX} .
Here, we define the family of the non-Hermitian Hamiltonian as
\begin{align}
    h_{\lambda}(k) = \lambda h_{\mathrm{asym}}(k) + (1-\lambda) h_{\mathrm{FK}}(k).
\end{align}
Here, $h_{\mathrm{FK}}$ and $h_{\mathrm{asym}}$ are the Bloch Hamiltonian of the Falicov-Kimball model with a complex-valued interaction and the asymmetric hopping Hamiltonian, which are explicitly written as 
\begin{align}
    h_{\mathrm{FK}}(k) = b_{2}(k){\sigma}_{2}+b_{3}(k){\sigma}_{3}+(U-i\gamma)
    \left(
    \begin{array}{cc}
       0  & 0 \\
        0 & 1
    \end{array}
    \right),
    \label{eq:effFK-Bloch-Hamiltonian}
\end{align}
and
\begin{align}
    h_{\mathrm{asym}}=
   (v_1 +v_2 e^{ik})\sigma_1 \hspace{5mm} (v_1,v_2 > 0)
\end{align}
respectively.
The function $b_{2}(k)$ and $b_{3}(k)$ 
are given in Eqs.~\eqref{eq:b2} and \eqref{eq:b3} in the main text.
Since the asymmetric hopping Hamiltonian $h_{\mathrm{asym}}$ is expressed as
\begin{align}
\label{eq:asym}
    H_{\mathrm{asym}} &= v_1 \sum\limits_{j}(
        c^{\dagger}_{j\mathrm{A}}c_{j\mathrm{B}} + \mathrm{h.c.} \nonumber 
    ) \\
    &+ 
    v_2 \sum\limits_{j}(
        c^{\dagger}_{j+1\mathrm{A}}c_{j\mathrm{B}} + c^{\dagger}_{j+1\mathrm{B}}c_{j\mathrm{A}}
    ) 
\end{align}
in real space, the second term denotes the asymmetric hopping from left to right, which leads to the localization of the eigenstate at the right edge.
The schematic illustration of the Hamiltonian is given in Fig.~\ref{fig:asym}.
\begin{figure}[!h]
\centering
    \includegraphics[width=1\hsize,clip]{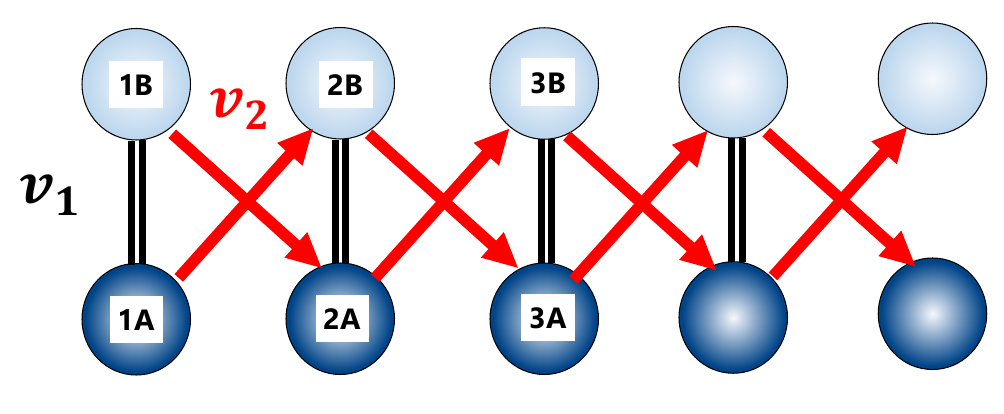}
\caption{Schematic figure of the Hamiltonian given in Eq.~\eqref{eq:asym}}
\label{fig:asym}
\end{figure}
Now we show that there exists a path to connect the Hamiltonians $h_{\mathrm{FK}}$ and $h_{\mathrm{asym}}$ without point-gap closing.
\begin{figure}[H]
\centering
    \includegraphics[width=1\hsize,clip]{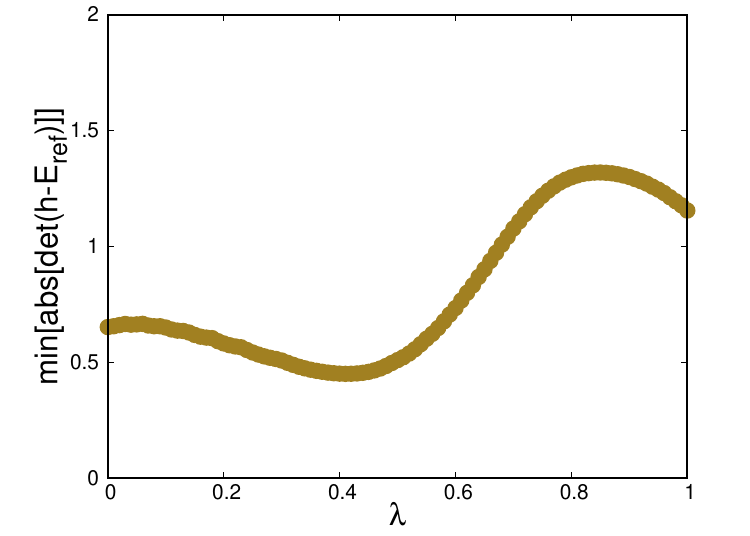}
\caption{The minimum value of $\abs{\det[h_{\mathrm{\lambda}}(k)-E_{\mathrm{ref}}]}$ as a function of $\lambda$.
By varying the 
The parameters are set to be $v_1 = 2,~v_2 = 1,~t=1,~U=0.1,~\gamma=1.0, E_{\mathrm{ref}}=-2 - 0.5i$.}
\label{fig:determinant}
\end{figure}
Figure~\ref{fig:determinant} shows the minimum value of the 
\begin{align}
    \abs{\det[h_{\mathrm{\lambda}}(k)-E_{\mathrm{ref}}]}
\end{align}
by varying the $\lambda$ from $0$ to $1$.
Remarkably, the point-gap at $E_{\mathrm{ref}}$ is always open, because $\abs{\det[h_{\mathrm{\lambda}}-E_{\mathrm{ref}}]}$ does not take the zero for any $\lambda$.
Thus, the non-Hermitian Hamiltonian $h_{\mathrm{asym}}$ is topologically equivalent to the Falicov-Kimball Hamiltonian $h_{\mathrm{FK}}$. It worth noting that by changing $k \longleftrightarrow -k$ in Eq.~\eqref{eq:effFK-Bloch-Hamiltonian}, the eigenstate is localized on the left side. 
}

\bibliography{RedEP}

\end{document}